\documentclass[10pt]{article}
\usepackage{bm}
\usepackage{amssymb}
\usepackage{graphicx}
\usepackage{algorithm}
\usepackage{algorithmic}
\usepackage{amsmath}
\usepackage{pict2e}
\usepackage{caption}
\usepackage{afterpage}

\def\shorttitletext{Reversible Elastic Collisions}
\def\titletext{Reversible Simulations of Elastic Collisions}

\floatname{algorithm}{Procedure}
\newlength{\myimagewidth}
\newlength{\myimageheight}
\newenvironment{myitemize}{%
    \begin{list}{$\bullet$}{\setlength{\leftmargin}{0.5in}}%
}{%
    \end{list}
}
\newcommand{\rotatedleadsto}[1]{%
    \mathrel{\raisebox{.1em}{%
        \reflectbox{\rotatebox[origin=c]{#1}{$\leadsto$}}}}}

\newcommand{\remove}[1]{}
\newcommand{\refeqn}[1]{Equation~(\ref{eqn:#1})}
\newcommand{\reffig}[1]{Figure~\ref{fig:#1}}
\newcommand{\refsec}[1]{Section~\ref{sec:#1}}
\newcommand{\refappsec}[1]{Appendix~\ref{sec:#1}}
\newcommand{\refalgo}[1]{Procedure~\ref{algo:#1}}

\newcommand{\reftab}[1]{Table~\ref{tab:#1}}
\newcommand{\myvect}[1]{\vec{{#1}}}

\begin{document}

\markboth{K. S. Perumalla and V. A. Protopopescu}{\shorttitletext}

\title{\titletext\footnote{Association for Computing Machinery Transactions on Modeling and Computer Simulation (ACM TOMACS) Vol. 23, No. 2, 2013}}
\author{
  Kalyan S. Perumalla\thanks{\texttt{perumallaks@ornl.gov}} \qquad Vladimir A. Protopopescu\\
    \textit{Oak Ridge National Laboratory}\\
    \textit{One Bethel Valley Rd, Oak Ridge, TN 37831-6085, USA}
}

\maketitle

\begin{abstract}
Consider a system of $N$ identical hard spherical particles moving in a
$d$-dimensional
box and undergoing elastic, possibly multi-particle, collisions.  We develop a
new algorithm that recovers the pre-collision state from the post-collision
state of the system, across a series of consecutive collisions, \textit{with
essentially no memory overhead}.
The challenge in achieving reversibility for an $n$-particle
collision (where, in general, $n\ll N$) arises from the presence of $nd-d-1$
degrees of freedom (arbitrary angles) during each collision, as well as from
the complex geometrical constraints placed on the colliding particles.  To
reverse the collisions in a traditional simulation setting, all of the
particular realizations of these degrees of freedom (angles) during the forward
simulation must be tracked.  This requires memory proportional to the number of
collisions, which grows very fast with $N$ and $d$, thereby severely limiting
the \textit{de facto} applicability of the scheme.  This limitation is
addressed here by first performing a pseudo-randomization of angles, which
ensures determinism in the reverse path for any values of $n$ and $d$.  To
address the more difficult problem of geometrical and dynamic constraints, a
new approach is developed which correctly samples the constrained phase space.
Upon combining the pseudo-randomization with correct phase space sampling,
perfect reversibility of collisions is achieved, as illustrated for $n\leq 3$,
$d=2$, and $n=2$, $d=3$.  This result enables, for the first time, reversible
simulations of elastic collisions with essentially zero memory accumulation.
In principle, the approach presented here could be generalized to larger
values of $n$.  The reverse computation methodology presented here uncovers
important issues of irreversibility in conventional models, and the difficulties
encountered in arriving at a reversible model for one of the most basic and
widely used physical system processes, namely, elastic collisions for hard
spheres.  Insights and solution
methodologies, with regard to accurate phase space coverage with reversible
random sampling proposed in this context, can help serve as models
and/or starting points for other reversible simulations.
\end{abstract}

\clearpage
\tableofcontents
\clearpage

\section{Introduction}\label{sec:intro}

\subsection{Background}

Modeling and simulation of particle collisions for a wide range of collision
types has been a subject of scientific interest for a long time.
However, while the \textit{forward} simulation of collisions
is relatively well understood,
their \textit{reversible} simulation is much less so.
Here, we study the problem of modeling and simulating elastic collisions
\textit{reversibly}.

In a system of $N$ identical hard spherical particles colliding with each other in a
$d$-dimensional box, every elastic collision, in general, yields an
underspecified system of equations and inequalities \cite{Truesdell1980}.  The
underspecified nature of the system gives rise to interesting challenges for
reversibility, such as the issue of memory accumulation and the need for
unbiased phase space coverage.

In each multi-particle collision of $n\ll N$ hard particles in $d$ dimensions,
the velocities represent $nd$ variables in the post-collision state, which are
related to the $nd$ velocities in the pre-collision state via conservation laws.
Upon applying conservation of momenta and kinetic energy, one is left with
$nd-d-1$ unspecified degrees of freedom.
\textit{Deterministic} solutions to this
underspecified system add new constraints that correspond to
specific additional assumptions on
the collisions.  For example, in a 2-particle collision,
determinism may be induced by exchanging velocity components
along the line joining the centers of the two
particles \cite{Lubachevsky1991,Marin1997PADS,Marin1997CPC,Hontalas1989}.
\textit{Non-deterministic} solutions can be
obtained by infusing the appropriate amount of randomization needed to
cover the phase space of the underspecified system
in a complete and unbiased manner.  Of particular concern,
is accounting for geometric constraints, \textit{e.g.}, disallowing
particle-overlapping at all times.

The goal of our work is to develop a modeling
and simulation framework which accurately and completely recovers the
pre-collision state from the post-collision state of all particles involved in
every collision in a sequence of collisions in the system,
\textit{with essentially no memory overhead}.

\subsection{Organization}

In \refsec{probdef}, the problem is defined and terminology is introduced.
The basic outline of our approach to reversal of
elastic collisions is outlined in \refsec{perfectsol}.  This is followed by
\refsec{algoimpl} which gives detailed
reversal algorithms for 2-particle collisions (up to 3 dimensions)
and 3-particle collisions (up to 2 dimensions).
Results from experiments with a software realization of the algorithms
are presented in \refsec{results}.
An estimation of the potential performance gains from reverse computation,
in comparison to conventional state saving approaches, is given in
\refsec{performance}.
The findings are summarized in \refsec{summary}.

\section{Problem Definition and Terminology}\label{sec:probdef}

\subsection{Reversibility Problem}

Consider a system consisting
of $N$ identical hard spheres of diameter $D$ in a $d$ dimensional cubic box
undergoing elastic collisions among themselves and/or with the walls of the box.
Reversible simulation of particle collisions in the system means that,
at any moment, the simulation can be stopped and executed
backwards to arbitrary points in the past, potentially all the way to the
initial state, such that the positions and velocities of
particles are restored to the same values that the particles had at the
chosen point in the past.
\textit{The problem is to achieve this reversible
simulation with minimal or, ideally, no memory overhead}.
The reversal scheme must support starting the system with arbitrary initial
configurations,
and must be able to evolve the system to arbitrary numbers of collisions into
the future.  Moreover, given that
particles are marked with identifiers from 1 to $N$, the system state must be
restored to the correct initial identifier assignation.

An important requirement of the collision model is the uniform (unbiased)
coverage of all available phase space.  In particular, the scattering law
upon each
collision must uniformly sample the full range of restitution angles available
to each pair of particles upon their collision.

\subsection{Collision Configurations and Constraints}

Here we consider collisions of three types:
(1) single particle-wall collisions,
(2) $n$-particle-wall collisions, and
(3) $n$-particle collisions ($n>1$).
The first type is straightforward to reverse.
The second type is treated, without loss of generality, as a simultaneous
set of individual wall-particle collisions, followed by $n$-particle collision
(for any collisions that remain after the application of individual
wall-particle collisions).
The third is the more complex problem treated in the remainder of the paper.

The collision of a particle with a wall is modeled by changing the sign of the
velocity component that is orthogonal to the wall.
If a particle touches more than one wall simultaneously at an edge or
a corner of the box, all the appropriate
velocity components change their sign.  Other commonly used
deterministic boundary conditions such as periodic wall boundaries
\cite{Lubachevsky1991,Marin1997PADS,Marin1997CPC,Krantz1996}
can be reversed analogously.

For completeness, we list here the set of system configurations that are
either trivial configurations that are not of interest, or are simple
to solve separately without needing the complexity of our algorithm.
\begin{itemize}
\item Trivial or degenerate situations, in which particles never come in contact.
      Examples include all particles being at rest (zero kinetic energy), which
      is clearly an uninteresting case.
\item Situations in which particles do come in contact, but do not exchange momentum.
      Examples in the first category include initial conditions in which:
      \begin{itemize}
        \item all particles are at rest, or
        \item all $N$ particles have velocities along one
          coordinate axis (say $x$) and the area of the particles' projections
          on the $(y,z)$ plane is $\tfrac{\pi N D^2}{4}$ (this perpetually collides
          particles with walls but never results in any particle-particle collisions).
      \end{itemize}
      Examples in the second category include grazing collisions, for which
      the line passing through the particles' centers at the encounter time is
      perpendicular to their (parallel or anti-parallel) velocities.  This
      condition is easily detected and either ignored or a
      degenerate collision operator can be defined separately and applied.
\end{itemize}


\subsection{Dynamics and Geometry}


In an $n$-particle collision, let
$\myvect{V'}_i$ be the pre-collision velocity of particle $i$, and
$\myvect{V}_i$ its post-collision velocity.  For every pair of
particles $i$ and $j$ that
are in contact in the collision, let $\myvect{r}_{ji}$ be
the vector from center of particle $j$ to that of particle $i$ at
the collision moment.  Let the total momentum of the $n$ colliding
particles be $\myvect{M}$ and their total kinetic
energy be $E$.  Then, the dynamics and geometry require the following:

\begin{equation}
      \left.
  \begin{aligned}
    \sum_{i=1}^n{\myvect{V'}_i}   =\sum_{i=1}^n{{\myvect{V}}_i}&=\myvect{M}\\
    \sum_{i=1}^n{(\myvect{V'}_i)^2} =\sum_{i=1}^n{{(\myvect{V}_i)}^2}&=E>0\\
  \end{aligned}
      \right\}
  \text{Dynamics},
  \label{eqn:generaldynamics}
\end{equation}

\begin{equation}
  \begin{aligned}
    \forall{i,j}\,\text{such that particles}\\
    \text{$i$ and $j$ are in contact}
  \end{aligned}
    \left|
  \begin{aligned}
    \myvect{r}_{ji}\cdot(\myvect{V'}_i-\myvect{V'}_j)   &< 0 \,\text{(pre-collision)}\\
    \myvect{r}_{ji}\cdot(\myvect{V}_i-\myvect{V}_j) &> 0 \,\text{(post-collision)}\\
  \end{aligned}
      \right\}
  \text{Geometry}.
  \label{eqn:generalgeometry}
\end{equation}

In \refeqn{generalgeometry}, the strictness of the inequalities ($<$
instead of $\leq$, and $>$ instead of $\geq$) ensures that
grazing collisions are excluded.

Let $d_n=nd-d-1$ denote the number of degrees of freedom in the collision.

Denote by $\Phi=\{\phi_k|1\leq k\leq d_n\}$ the set of ``parameters''
that, given $\myvect{M}$ and $E$, uniquely determines the set of velocities
${\bm V}=\{\myvect{V}_i| 1\leq i\leq n\}$ of all particles in an $n$-particle
collision.  Let $\Phi^{'}$ correspond to the parameters encoding the
pre-collision velocities, and let $\Phi$ (or, where ambiguous, $\Phi^{''}$)
encode the post-collision velocities.  In 2-particle collisions, the
parameters correspond to geometrical angles of points on the surface
of a $d_n$-sphere, whereas, in 3-particle collisions, the parameters
are different from the geometrical angles of points on the surface of
a $(d_n+1)$-dimensional ellipsoid.

In randomly sampling the phase space spanned by the $d_n$ parameters,
at least $d_n$ random numbers must be generated per collision.  Reversible
random number generators are used to generate the needed random samples per
collision, each sample value uniformly distributed in $[0,1)$.  The generators
themselves require only a constant amount of memory.  In practice, the memory
per generator is often only a few bytes long.  In theory, the memory
only needs to be independent of the number of collisions being
simulated.  Additional considerations in random number generation that are
critical to reversibility are discussed in \refsec{rngstreams}.

Our reversible collision algorithm uses the
following mappings.

\textbf{${\bm V}$-to-${\bm \Phi}$}:
Given ${\bm V}$, this mapping determines the set of angles $\Phi$ that uniquely
determines ${\bm V}$.  In forward execution, this mapping will be applied on
pre-collision velocities ${\bm V'}$ to obtain $\Phi^{'}$.  In reverse
execution, this mapping function will be used to determine $\Phi$ from the
post-collision velocities ${\bm V}$ in a collision.

\textbf{${\bm \Phi}$-to-${\bm V}$}:
Given  the angles $\Phi$, together with $\myvect{M}$ and $E$,
this mapping reconstructs all corresponding ${\bm V}$.  In forward
execution, this mapping will be used to generate the post-collision velocities
${\bm V}$ based on reversibly computed values of $\Phi$.  The same mapping
function will also be used in reverse execution to determine the pre-collision
velocities ${\bm V'}$ from the recovered value of $\Phi^{'}$ of a collision.

\textbf{${\bm G}$}:
Let ${\bm G}=\{G_k|1\leq k\leq d_n\}$ denote the set of pseudo random
numbers generated for each collision, where each $G_k\in[0,1)$.  The generator
used to generate ${\bm G}$ is reversible, of high quality, and
with a sufficiently long period.  The generator can either be a single stream
or contain $d_n$ independent, parallel streams.

\textbf{${\bm G}$-to-${\bm \Psi}$}:
Further, let $\Psi=\{\psi_k|1\leq k\leq d_n\}$, be the set of angle
\textit{offsets} generated from random numbers ${\bm G}$.
In both forward as well as reverse execution, the \textbf{${\bm G}$-to-${\bm
\Psi}$} function will be used as a deterministic mapping from uniform random
numbers to angle offsets.
The specific mapping function depends on $n$ and $d$, but
the function only depends on the (reversible) random numbers, total momenta and
total energy; it does not depend on the individual velocities of particles in
collision.

In the remainder of the article, unless otherwise specified, all arithmetic on
the angles will be performed modulo $2\pi$.

\subsection{Simplified Notation for $2\leq n\leq 3$, $1\leq d\leq 3$}

When dealing with collisions in which at most three particles are in contact
with each other, in 1- or 2-dimensional space (and also in the case in
which two particles are in contact with each other in 3-dimensional space),
a simplified notation is used in the remaining of the article to refer to their
velocities, momenta, and energy.  The letters $a$--$f$ will be used to refer to
the components of the velocities along the $x$, $y$, and $z$ directions.  The
letters $a'$--$f'$ will be used to refer to the corresponding pre-collision
values of the velocity components.  The letters $\alpha$, $\beta$, and $\gamma$
will be used as the sums of momenta along the $x$, $y$ and $z$ spatial
directions respectively, and $\delta$ will be used for total kinetic energy.
Thus, for example, in a 2-dimensional, 2-particle collision, the post-collision
equations will be written as $a+b=\alpha$ (total momentum in the
$x$-direction), $c+d=\beta$ (total momentum in the $y$-direction), and
$a^2+b^2+c^2+d^2=\delta$ (total energy), where $a={V_{1x}}$, $b={V_{2x}}$,
$c={V_{1y}}$ and $d={V_{2y}}$ are the velocities of the two particles in the
$x$ and $y$ directions.
In the remainder of the article, note that $\delta>0$, since the system must
have non-zero kinetic energy for collisions to occur.

\section{Skeleton of the Algorithm}\label{sec:perfectsol}

In any $n$-particle collision, the information available at hand during forward
execution are the pre-collision velocities as well as the ranges of
the angles that determine the range of permissible post-collision
velocities.  The pre-collision velocity configuration of all the $n$ colliding
particles can be uniquely encoded in terms of the total momentum, total energy,
and the specific values of the free angles.
Since all collisions are elastic, the total momentum and energy
do not need any memory to recover.  The problem thus reduces to that of
developing a one-to-one mapping of pre-collision and post-collision
angles, while still uniformly sampling all available phase space for
the angles at every collision.

In order to properly sample the available phase space, we use pseudo random
numbers to select the angle values from the permissible ranges.
The reversibility of the pseudo random number generators ensures the ability
to go forward as well as backward in the random number sequence as
needed, during forward and reverse execution, respectively.

\subsection{Reversible Collision Operation}

The problem reduces now to developing a collision
algorithm that (a) takes $d_n$ pseudo random numbers, each uniformly
distributed in $[0,1)$, and gives $d_n$ reversible random angle offsets that
satisfy the pre-collision phase space constraints, and (b) recovers
the pre-collision angles from the random offsets recovered upon backward
execution.

Once such a collision algorithm is developed, it can be used to uniquely
recover the pre-collision velocities as follows.  First, the random number
sequence is reversed, thereby recovering the random numbers that were used in
the forward execution.  These are then used to recreate the random offsets
that were used in the forward execution.  The random offsets are applied
in the opposite direction on the post-collision values of the free angles
to uniquely recover the pre-collision values of the free angles, which in
turn uniquely give the pre-collision velocities.

\subsection{Collision Sequences}

Collision \textit{sequences} are modeled using standard
techniques \cite{Lubachevsky1991,Miller2004,Krantz1996},
by which, at every step of forward evolution,
the time for next collision of each particle is determined, time is advanced to
the earliest collision time, the particles undergoing collision are determined,
their pre-collision velocities are transformed by our reversible algorithm
to give post-collision velocities, and the process is repeated.

In order to reverse the collision \textit{sequence}, it is necessary to recover
the most recent collision event in the past.  That event is obtained by
reversing the direction of all particles' velocities and employing the usual
forward algorithm for determining the next earliest collision.  The event
represents information on the amount of time, $dt$, to go back in time, and the
identities of the colliding particles.  The system is then stepped back in time
by $dt$ units by changing the sign of the velocities of all particles, and
linearly transporting them for $dt$ units.  At that moment, clearly, the
colliding particles would be found to be in contact.  The reverse collision
algorithm is then applied on the particles in contact.  This process is
repeated iteratively until the time of interest in the past is reached.

\subsection{General and Specific Settings}

This general reversal scheme is applicable in principle to any values of $n$
and $d$.  However, the actual permissible ranges of the free angle values
remain to be determined for each $(n,d)$ pair.  As illustrations, we develop
the geometrical constraints for the subsets $n\leq 3, d \leq 2$ and $n=2, d=3$.
While the development of the permissible angle ranges for $n=2$ is relatively
straightforward, those for $n=3$ become complex starting even from $d=1$.

\section{Algorithm Implementation}\label{sec:algoimpl}

Here, we determine the phase space of the permissible reversible
random offsets that need to be sampled for 2-particle collisions in 1, 2, and 3
dimensions, and for 3 particles in 1 and 2 dimension.

\subsection{Reversal for 2-Particle Collisions in 1 Dimension}

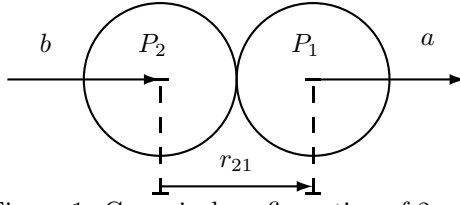
\begin{figure}[htbp]
\centering
\setlength{\unitlength}{0.8in}
\begin{picture}(3.5,1.25)(0,-0.1)
  \thicklines
  \put(0.5,0.5){\circle{1.0}}
  \put(-0.5,0.5){\vector(1,0){1.0}}
  \put(-0.25,0.8){\makebox(0,0)[t]{$b$}}
  \put(0.45,0.8){\makebox(0,0)[t]{$P_2$}}
  \put(1.5,0.5){\circle{1.0}}
  \put(1.5,0.5){\vector(1,0){1.0}}
  \put(2.25,0.8){\makebox(0,0)[t]{$a$}}
  \put(1.45,0.8){\makebox(0,0)[t]{$P_1$}}
  \put(0.5,-0.20){\vector(1,0){1.0}}
  \put(1.0,0.00){\makebox(0,0)[t]{$r_{21}$}}
  \put(0.5,-0.25){\dashbox{0.1}(0,0.75)}
  \put(1.5,-0.25){\dashbox{0.1}(0,0.75)}
\end{picture}
\caption{Canonical configuration of 2 particles in 1 dimension}
\label{fig:2spheres1dimconfig}
\end{figure}

In a collision of 2 particles in 1-dimensional space,
as shown in \reffig{2spheres1dimconfig},
let $P_1$ be the particle on the right moving with velocity $a$,
and $P_2$ be the particle on the left moving with velocity $b$.
The constraints on dynamics and post-collision geometry are:

\begin{equation}
      \left.
  \begin{aligned}
    a+b&=\alpha\\
    a^2+b^2&=\delta\text{, }
    2 \delta&> \alpha^2
  \end{aligned}
      \right\}
  \text{Dynamics},
\end{equation}

\begin{equation}
    \left.
  \begin{aligned}
    r_{21}\cdot(a-b)& > 0\\
    \text{for } r_{21}=D & > 0
  \end{aligned}
      \right\}
  \text{Geometry}.
\end{equation}

The pre-collision geometrical constraints are obtained
by replacing $> 0$ by $< 0$ in the post-collision constraints.
The system is fully defined without any free angles, giving
deterministic one-to-one mapping from pre-collision to post-collision
velocities and hence reversibility is unambiguous.  Nevertheless, this
configuration is treated here for completeness.
The equations of motion imply that:

\begin{equation}
    \left.
  \begin{aligned}
    a&=\frac{\alpha}{2}+\frac{\sqrt{2\delta-\alpha^2}}{2} \text{, and }
    b&=\frac{\alpha}{2}-\frac{\sqrt{2\delta-\alpha^2}}{2}
  \end{aligned}
      \right\}
  \text{Dynamics},
\end{equation}

\begin{equation}
    \left.
  \begin{aligned}
    r_{21}\cdot(a-b)&=+r_{21}\sqrt{2\delta-\alpha^2}>0
  \end{aligned}
      \right\}
  \text{Geometry},
\end{equation}

\noindent
\textit{i.e.}, the dynamical solution satisfies
the geometrical constraint.

\subsection{Reversal for 2-Particle Collisions in 2 Dimensions}\label{sec:2spheres-2dim}

In a collision of 2 particles in 2-dimensional space, let $a$ and $c$
be the $x$ and $y$ velocity components of the particle with the
smaller identifier, and $b$ and $d$ be the corresponding velocity components
of the other particle.  The equations of motion and post-collision geometry are:

\begin{equation}
      \left.
  \begin{aligned}
    a+b&=\alpha\text{, }
    c+d =\beta\\
    a^2+b^2+c^2+d^2&=\delta\text{, and }
    2\delta >\alpha^2+\beta^2
  \end{aligned}
      \right\}
  \text{Dynamics},
\end{equation}

\begin{equation}\label{eqn:2sphere-2dim-geometry-constraint}
    \left.
  \begin{aligned}
    r_{21x}\cdot(a-b)+r_{21y}\cdot(c-d)& > 0\\
    \text{for some } r_{21x},r_{21y} & > 0
  \end{aligned}
      \right\}
  \text{Geometry}.
\end{equation}

The pre-collision geometrical constraints are obtained
by replacing $> 0$ by $< 0$ in the post-collision constraints.
The equations of motion imply that:

\begin{equation}
  \begin{aligned}
    \left( a-\frac{\alpha}{2} \right)^2 +
    \left( c-\frac{\beta}{2} \right)^2
    &=\frac{2\delta-(\alpha^2+\beta^2)}{4}=R^2,
  \end{aligned}
\end{equation}

\noindent
\textit{i.e.}, the point ($a$, $c$) lies on a circle of radius
$R=\sqrt{2\delta-(\alpha^2+\beta^2+\gamma^2)}/2$ centered at
($\alpha/2$, $\beta/2$), whose parametric equations are given by:

\begin{equation}\label{eqn:2sphere-2dim-geometry}
  \begin{aligned}
    a&=\frac{\alpha}{2}+R\cos{\phi_1}\text{ , }
    c&=\frac{\beta}{2}+R\sin{\phi_1}\text{ , and }
    \phi_1&\in[0,2\pi).\\
  \end{aligned}
\end{equation}
\remove{
satisfies $r_{21x}\cdot(a-b)+r_{21y}\cdot(c-d) \geq 0$ for
pre-collision velocities, and $r_{21x}\cdot(a-b)+r_{21y}\cdot(c-d) \leq 0$ for
post-collision velocities
}
\refeqn{2sphere-2dim-geometry} provides the
\textbf{${\bm V}$-to-${\bm \Phi}$} and
\textbf{${\bm \Phi}$-to-${\bm V}$}
mapping functions for 2-dimensional 2-particle collisions.
Given $(a',b',c',d')$, a unique $\phi_1^{'}$ can be determined.  Similarly,
given $(\phi_1,\alpha,\beta,\delta)$, $(a,b,c,d)$ can be uniquely determined.

\begin{figure}[htbp]
  \centering
  \includegraphics[width=0.95\myimagewidth,keepaspectratio]{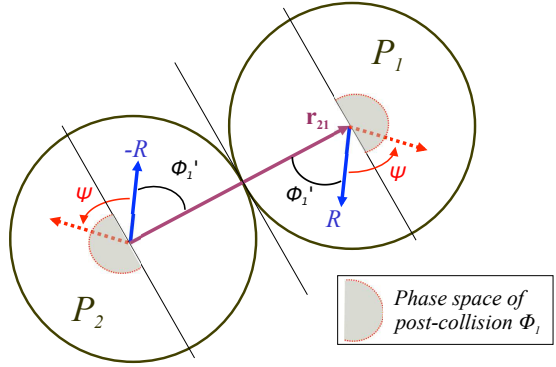}
  \caption{Configuration of 2 particle collision in 2 dimensions in the center-of-mass frame}
  \label{fig:2part2dim-comass-fwd}
\end{figure}

Given the pre-collision angle $\phi_1^{'}$, the problem at hand is the
generation of a reversible random offset $\psi$ from $\phi_1^{'}$ to give the
post-collision angle $\phi_1=\phi_1^{'}+\psi$ which can then be used to
determine the post-collision velocities.
Since the phase space of $\phi_1$ is $[0,2\pi)$, it is necessary to sample the
random offset $\psi$ also from the same range, in order to
ensure full phase space coverage independent of $\phi_1^{'}$.  This is illustrated
in \reffig{2part2dim-comass-fwd} in which the colliding pair is visualized in its
center-of-mass frame of reference (in this particular case of 2-particles in
2-dimensions, the geometric angle $\phi_1$ indeed corresponds to the degree of
freedom, but this does is not true in general).
Moreover since the circumference of the circle is uniformly sampled by uniformly
sampling the angle subtended at the center, it is sufficient to
generate $\psi$ uniformly from 0 to $2\pi$.
Thus, $\psi\in [0,2\pi)$ is generated with a
\textbf{${\bm G}$-to-${\bm \Psi}$} mapping given by
$\psi=2G\pi$, with a uniformly distributed random number $G\in[0,1)$.

However, the $\phi_1$ value thus computed may violate the geometry
constraint \refeqn{2sphere-2dim-geometry-constraint}
on post-collision velocities.
If such a violation occurs with
the generated random offset, the $\phi_1$ value can be ``corrected'' by adding
$\pi$ to it; the addition of $\pi$ guarantees satisfaction of the geometry
constraint on post-collision velocities because of the following observations:

\begin{equation}\label{eqn:2sphere-2dim-phirelations}
  \begin{aligned}
    \text{Since } & (a-b)=2R\cos\phi_1 \text{, and } (c-d)=2R\sin\phi_1,\\
    \text{if } & r_{21x}\cdot(a-b)+r_{21y}\cdot(c-d) & < 0\\
    \text{then } & r_{21x}\cdot(2R\cos\phi_1)+r_{21y}\cdot(2R\sin\phi_1) & < 0\\
    \implies & r_{21x}\cdot(2R\cos(\phi_1+\pi))+r_{21y}\cdot(2R\sin(\phi_1+\pi)) & > 0 .\\
  \end{aligned}
\end{equation}

Thus, the post-collision angle, using the random angle offset
$\psi$, is computed as either ${\bm F:}~\phi_1=\phi_1^{'}+\psi$
or ${\bm F_\pi:}~\phi_1=\phi_1^{'}+\psi+\pi$.
In reverse execution, the pre-collision
angle $\phi_1^{'}$ is recovered as ${\bm R:}~\phi_1^{'}=\phi_1-\psi$ or
${\bm R_\pi:}~\phi_1^{'}=\phi_1-\psi-\pi$.  If the fact that
$\pi$ was added by the forward execution is somehow remembered (\textit{i.e.},
whether ${\bm F}$ or ${\bm F_\pi}$ was applied), then
the correct pre-collision angle can be recovered during reversal
(by applying ${\bm R}$ or ${\bm R_\pi}$ respectively).

Our key observation here is that it is possible to avoid having to
``remember'' whether the $\pi$ offset was added or not.
If $\pi$ was added in forward execution, then, in reverse execution,
the fact that $\psi+\pi$ must be used, as opposed to
$\psi$, can be detected by a violation of the geometry constraint on the
pre-collision velocities if $\pi$ is not added.
The correct value of $\phi_1^{'}=\phi_1-\psi-\pi$ is thus
possible to be recovered correctly.

To adopt this approach, it is necessary and sufficient
to show that no ambiguity exists during the reverse execution regarding
which of ${\bm F}$ or ${\bm F_\pi}$ was executed in the forward execution,
so that the correct operation, ${\bm R}$ or ${\bm R_\pi}$, is applied
for correct reversal.  In other words, it is necessary and sufficient
to satisfy the following conditions:\\
\begin{tabular}{p{0.31\textwidth}p{0.31\textwidth}p{0.31\textwidth}}
\begin{myitemize}
  \item ${\bm R}({\bm F}(\phi))=\phi$
  \item ${\bm R_{\pi}}({\bm F}(\phi))\neq\phi$
\end{myitemize}
&
\begin{myitemize}
  \item ${\bm R_{\pi}}({\bm F_{\pi}}(\phi))=\phi$
  \item ${\bm R}({\bm F_{\pi}}(\phi))\neq\phi$
\end{myitemize}
&
\begin{myitemize}
  \item ${\bm F}(\phi)\neq {\bm F_{\pi}}(\phi))$
  \item ${\bm R}(\phi)\neq {\bm R_{\pi}}(\phi))$.
\end{myitemize}
\end{tabular}

These conditions can be proven as follows.
Let $CS(\phi)=r_{21x}\cdot(2R\cos\phi)+r_{21y}\cdot(2R\sin\phi)$.
We know that
 ${\bm F}$ and ${\bm F_\pi}$ are exclusive with respect to satisfaction
 of \refeqn{2sphere-2dim-geometry-constraint} with $\phi_1$, \textit{i.e.},
  ${\bm F}$ satisfies $CS(\phi_1) > 0$,
    if and only if ${\bm F_\pi}$ satisfies $CS(\phi_1) < 0$.
Similarly,
 ${\bm R}$ and ${\bm R_\pi}$ are exclusive with respect to satisfaction
 of \refeqn{2sphere-2dim-geometry-constraint} with $\phi_1^{'}$, \textit{i.e.},
  ${\bm R}$ satisfies $CS(\phi_1^{'}) < 0$,
    if and only if ${\bm R_\pi}$ satisfies $CS(\phi_1^{'}) > 0$.

The geometrical constraints require
that the post-collision velocities satisfy $CS(\phi_1) > 0$ and the
pre-collision velocities satisfy $CS(\phi_1^{'}) < 0$.  These requirements
are satisfied if ${\bm R}$ is used to reverse ${\bm F}$, or ${\bm R_\pi}$
is used to reverse ${\bm F_\pi}$, respectively.  In other words, from
\refeqn{2sphere-2dim-phirelations}, we know that:\\
\indent$\bullet$ if ${\bm F}$ satisfies $CS(\phi_1) > 0$ in forward,
    then ${\bm R}$ satisfies $CS(\phi_1^{'}) < 0$ in reverse, and\\
\indent$\bullet$ if ${\bm F_\pi}$ satisfies $CS(\phi_1) > 0$ in forward,
    then ${\bm R_\pi}$ satisfies $CS(\phi_1^{'}) < 0$ in reverse.

Finally, the ambiguity is fully resolved when the following are also satisfied:\\
\indent$\bullet$  If ${\bm F}$ satisfies $CS(\phi_1) > 0$,
    then ${\bm R_\pi}$ violates $CS(\phi_1^{'}) < 0$.\\
\indent$\bullet$  If ${\bm F_\pi}$ satisfies $CS(\phi_1) > 0$,
    then ${\bm R}$ violates $CS(\phi_1^{'}) < 0$.

The former can be proved as follows, and the latter can be proved along
similar lines:
Since $CS(\phi_1^{'}) < 0$ (pre-collision), and ${\bm F}$ gives $CS(\phi_1) > 0$ (post-collision),
${\bm R_\pi}$ gives $CS(\phi_1-\psi-\pi)$
    $= CS(\overline{\phi_1^{'}+\psi}-\psi-\pi)$
    $= CS(\phi_1^{'}-\pi)$
    $= -CS(\phi_1^{'})$.
This implies $CS(\phi_1^{'})  = -CS(\phi_1^{'})$, which is a contradiction.

The forward and reverse algorithms are given in \refalgo{2part2dimforward}
and \refalgo{2part2dimreverse} respectively.
This completes the generation of reversible random offsets for all
configurations of 2-particle collisions in 2 dimensions, ensuring
full phase space coverage with zero memory overhead.

\begin{algorithm}[phtb]
    \caption{($\phi_1^{'} \longrightarrow \phi_1$): \textit{Forward Function for 2-Particle in 2 Dimensions}}
    \label{algo:2part2dimforward}
    \begin{algorithmic}[1]
        \STATE $\psi \leftarrow 2G\pi$ \COMMENT{generate a random offset from $[0,2\pi)$}
        \STATE $\phi_1 \leftarrow (\phi_1^{'}+\psi) \mod 2\pi$ \COMMENT{post-collision is pre-collision offset by randomized $\psi$}
        \STATE \COMMENT{Next, if post-collision is converging, correct it to be diverging}
	\IF {$r_{21x}\cdot\cos{\phi_1}+r_{21y}\cdot\sin{\phi_1} < 0$}
            \STATE $\phi_1 \leftarrow (\phi_1+\pi) \mod 2\pi$
        \ENDIF
    \end{algorithmic}
\end{algorithm}
\begin{algorithm}[phtb]
    \caption{($\phi_1 \longrightarrow \phi_1^{'}$): \textit{Reverse Function for 2-Particle in 2 Dimensions}}
    \label{algo:2part2dimreverse}
    \begin{algorithmic}[1]
      \STATE $\psi \leftarrow 2G\pi$ \COMMENT{recover the random offset}
      \STATE $\phi_1^{'} \leftarrow (\phi_1-\psi) \mod 2\pi$ \COMMENT{initial guess at pre-collision angle}
      \STATE \COMMENT{Next, if pre-collision is diverging, correct it to be converging}
      \IF {$r_{21x}\cdot\cos{\phi_1}+r_{21y}\cdot\sin{\phi_1} > 0$}
          \STATE $\phi_1^{'} \leftarrow (\phi_1^{'}-\pi) \mod 2\pi$
      \ENDIF
    \end{algorithmic}
\end{algorithm}

\subsection{Reversal for 2-Particle Collisions in 3 Dimensions}\label{sec:2spheres-3dim}

In a collision of 2 particles in 3-dimensional space, the
equations of motion and post-collision geometry are:

\begin{equation}
      \left.
  \begin{aligned}
    a+b =\alpha\text{, }
    c+d =\beta\text{, }
    e+f =\gamma\\
    a^2+b^2+c^2+d^2+e^2+f^2 =\delta\text{, and }
    2\delta >\alpha^2+\beta^2+\gamma^2
  \end{aligned}
      \right\}
  \text{Dynamics},
\end{equation}

\begin{equation}\label{eqn:2sphere-3dim-geometry-constraint}
    \left.
  \begin{aligned}
    r_{21x}\cdot(a-b)+r_{21y}\cdot(c-d)+r_{21z}\cdot(e-f) & > 0\\
    \text{for some } r_{21x},r_{21y},r_{21z} & > 0 &
  \end{aligned}
      \right\}
  \text{Geometry}.
\end{equation}

The pre-collision geometrical constraints are obtained
by replacing $> 0$ by $< 0$ in the post-collision constraints.
The equations of motion imply that:

\begin{equation}
  \begin{aligned}
    \left( a-\frac{\alpha}{2} \right)^2 +
    \left( c-\frac{\beta}{2} \right)^2 +
    \left( e-\frac{\gamma}{2} \right)^2
    =\frac{2\delta-(\alpha^2+\beta^2+\gamma^2)}{4}=R^2,
  \end{aligned}
\end{equation}

\noindent
\textit{i.e.}, the point ($a$, $c$, $e$) lies on a sphere of radius
$R=\tfrac{1}{2}\sqrt{2\delta-(\alpha^2+\beta^2+\gamma^2)}$ centered
at ($\tfrac{\alpha}{2}$, $\tfrac{\beta}{2}$, $\tfrac{\gamma}{2}$).
whose parametric equations are given by:

\begin{equation}\label{eqn:2sphere-3dim-geometry}
  \begin{aligned}
    a =\frac{\alpha}{2}+R\sin{\phi_1}\sin{\phi_2}\text{, }
    c =\frac{\beta}{2}+R\sin{\phi_1}\cos{\phi_2}\text{, }
    e =\frac{\gamma}{2}+R\cos{\phi_1}\text{,}\\
    \phi_1 \in[0,\pi]\text{, and }
    \phi_2 \in[0,2\pi)~.
  \end{aligned}
\end{equation}
\refeqn{2sphere-3dim-geometry} provides the
\textbf{${\bm V}$-to-${\bm \Phi}$} and
\textbf{${\bm \Phi}$-to-${\bm V}$}
mapping functions for 2-particle collisions in 3 dimensions.

In determining ${\bm \Phi}$ from ${\bm V}$, if $\phi_1=0$ or $\phi_1=\pi$,
then, the value of $\phi_2$ is immaterial.  In that case,
we choose to set $\phi_2=0$.  Setting $\phi_2$ thus,
without regard to $\phi_2^{'}$, loses information about $\phi_2^{'}$ when
$\phi_1=0$.  To deal with this rare special case, the value of $\phi_2^{'}$
can be logged in the forward collision, and restored from the log
in the backward collision.  Note that this is logged in forward execution
\textit{only} if $\phi_1=0$, and \textit{not} for every collision.
In the reverse execution, $\phi_2^{'}$ is recovered from the log only
if $\phi_1=0$.

For the \textbf{${\bm G}$-to-${\bm \Psi}$} mapping, the offsets $\psi_1$ and
$\psi_2$ are obtained by uniformly sampling the surface of a unit sphere.  Any
algorithm for this purpose can be employed (e.g.,
\cite{Marsaglia1972}), using two random numbers $G_1\in[0,1]$ and
$G_2\in[0,1)$.

Similar to the 2-dimensional 2-particle case,
the post-collision angles are computed as
$\phi_1=(\phi_1^{'}+\psi_1) \mod 2\pi$ or
$\phi_1=(\phi_1^{'}+\psi_1+\pi) \mod 2\pi$,
and $\phi_2=(\phi_2^{'}+\psi_2) \mod 2\pi$.
The choice of whether $\pi$ is added to $\psi$ is
determined by the one that satisfies the geometrical condition given by
\refeqn{2sphere-3dim-geometry-constraint}.  The proof of reversibility of the
computation of $\phi_1$ is analogous to the one in the preceding sub-section.

This completes the generation of reversible random offsets for all
configurations of 2-particle collisions in three dimensions, ensuring
full phase space coverage with zero memory overhead.

\subsection{Reversal for 3-Particle Collisions in 1 Dimension}\label{sec:3spheres-1dim}

In this section, we determine the phase space of the permissible reversible
random offsets that needs to be sampled for 3-particle collisions in
1 dimension.

\subsubsection{Solving the Equations of Motion}

The equations of motion are:

\begin{equation}
      \left.
  \begin{aligned}
    a+b+c&=\alpha\\
    a^2+b^2+c^2&=\delta\text{, }
    3\delta&>\alpha^2
  \end{aligned}
      \right\}
  \text{Dynamics}.
\end{equation}

The post-collision geometrical constraints are given by:
\begin{equation}\label{eqn:3sphere-1dim-geometry-orig}
      \left[
  \begin{aligned}
    \text{Only two of these three need be}\\
    \text{satisfied for any given geometric}\\
    \text{configuration } r_{21},r_{32},r_{13}>0
  \end{aligned}
      \right]
      \left.
  \begin{aligned}
    r_{21}\cdot(a-b)&> 0,\\
    r_{32}\cdot(b-c)&> 0,\\
    r_{13}\cdot(c-a)&> 0
  \end{aligned}
      \right\}
  \text{Geometry}.
\end{equation}
In \refeqn{3sphere-1dim-geometry-orig}, when any two inequalities are satisfied,
the third one is resulting.  This is because the two inequalities that are
satisfied imply a specific ordering of the three particles along
the single dimension, which in turn automatically satisfies the
remaining third constraint.
The pre-collision geometrical constraints are obtained
by replacing $> 0$ by $< 0$ in the post-collision constraints.
The solution to the equations of motion satisfies:
%
\begin{equation}\label{eqn:3sphere-1dim-ellipse}
  \begin{aligned}
    \bar{a}^2 +
    {\left(\frac{\bar{b}-\frac{\sqrt{2}}{3}\alpha}{\dfrac{1}{\sqrt{3}}}\right)}^2 &=
        \delta-\frac{\alpha^2}{3} \text{, where }
    \bar{a}&=\frac{a-b}{\sqrt{2}}\text{, and }
    \bar{b}&=\frac{a+b}{\sqrt{2}},
  \end{aligned}
\end{equation}
%
which leads to the parametrization:
%
\begin{equation}
  \begin{aligned}
    \bar{a}&=\frac{\lambda}{\sqrt{2}}\cos{\phi_1}\text{, }
    \bar{b}&=\frac{\sqrt{2}}{3}\alpha+\frac{\lambda}{\sqrt{2}\sqrt{3}}\sin{\phi_1}\text{, }
    \lambda&=\sqrt{2}\sqrt{\delta-\frac{\alpha^2}{3}}\text{, and }
    \phi_1&\in[0,2\pi).\\
  \end{aligned}
\end{equation}
%
Converting to $a$, $b$ and $c$, we get:
\begin{eqnarray}\label{eqn:3sphere-1dim-geometry}
\begin{aligned}
a  = \frac{\alpha}{3}+\frac{\lambda}{2}(\cos{\phi_1}+\frac{1}{\sqrt{3}}\sin{\phi_1})\text{, }
b  = \frac{\alpha}{3}+\frac{\lambda}{2}(-\cos{\phi_1}+\frac{1}{\sqrt{3}}\sin{\phi_1})\text{,}\\
c  = \frac{\alpha}{3}+\frac{\lambda}{2}(-\frac{2}{\sqrt{3}}\sin{\phi_1})\text{, }
\lambda =\sqrt{2}\sqrt{\delta-\frac{\alpha^2}{3}}\text{, and }
\phi_1 \in[0,2\pi).
\end{aligned}
\end{eqnarray}
\refeqn{3sphere-1dim-geometry} provides the
\textbf{${\bm V}$-to-${\bm \Phi}$} and
\textbf{${\bm \Phi}$-to-${\bm V}$}
mapping functions for 1-dimensional 3-particle collisions.

\subsubsection{Sampling the ${\bm \Psi}$ Phase Space}

Analogously to the 2-particle 2-dimensional case, the phase space of the
post-collision angle is sampled by generating a random $\phi_1$ as a random
offset $\psi_1\in[0,2\pi)$ from the pre-collision angle $\phi_1^{'}$.  The
mapping function from a uniform random number $G\in[0,1)$ to $\psi_1$ is more
complex than that for the 2-particle case.  In the 2-particle case, the phase
space for velocities lies on the circumference of a circle, which can be
sampled simply by uniformly sampling the angle subtended at the center.
However, the phase space of the velocities in the 3-particle case lies on the
circumference of an ellipse, which cannot be sampled simply as $\psi_1=2G\pi$.
Instead, any correctly unbiased procedure for generating the angle by sampling
the circumference of an ellipse can be employed for the purpose of defining the
\textbf{${\bm G}$-to-${\bm \Psi}$} mapping function.  One such method is
described in the next section.

\subsubsection{Sampling the Circumference of the Ellipse}\label{sec:ellipsecircumf}

Here, we present a new procedure for uniformly sampling a point from the
perimeter of an ellipse.  The procedure is designed to be suitable for
use in reversible execution, requiring exactly one random number
per sample.  While rejection-based procedures exist for this problem,
they cannot be used here due to their irreversibility, as explained later
in \refsec{rngstreams}.

\begin{figure}[htbp]
  \centering
  \includegraphics[width=0.80\myimagewidth]{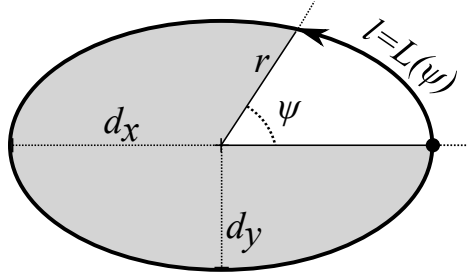}
  \caption{Sampling scheme for a uniformly selected random point on an ellipse}
  \label{fig:ellipse}
\end{figure}

Consider the ellipse
$\left(\tfrac{x}{d_x}\right)^2+\left(\tfrac{y}{d_y}\right)^2=1$ shown in
\reffig{ellipse}. Let
$L:\psi\rightarrow \ell$ be a function that maps any angle
\footnote{
We will use the terms angle and parameter interchangeably, with the
understanding that the angle is in fact the parameter in the representation
of the point on the ellipse, and \textit{not} the geometrical angle.}
$\psi$, $0\leq\psi<2\pi$, to
the length $\ell$ of the arc moving counter-clockwise along the
circumference of the ellipse from $(d_x,0)$ to $(r\cos{\psi},r\sin{\psi})$,
where $r^2={d_x^2 d_y^2}/({d_y^2\cos{\psi}^2+d_x^2\sin{\psi}^2})$.
Let $L^{-1}: \ell\rightarrow\psi$ be the inverse function that
determines the angle corresponding to any given arc length $\ell$.  Thus,
$L^{-1}(L(\psi))=\psi$.
A random point on the circumference of
the ellipse can be obtained as the end of the arc whose length is $G\cdot
L(2\pi)$, and the angle corresponding to that point can be obtained as
$\psi_1=L^{-1}(G\cdot L(2\pi))$, which serves as the \textbf{${\bm G}$-to-${\bm
\Psi}$} mapping for the 3-particle collisions in 1 dimension.

The value of $L(2\pi)$ is computable by different methods.  For example,
$L(2\pi)=\pi(d_x+d_y)\sum_{n=0}^{\infty}{\binom{0.5}{n}^2h^n}$, where
$h=\tfrac{(d_x-d_y)^2}{(d_x+d_y)^2}$, which can be computed to any desired
precision.  The value of $L^{-1}(\ell_{given})$ can be obtained by a bisection
method (binary search) that starts with the lowerbound $\psi_{lower}=0$
upperbound $\psi_{upper}=2\pi$, and an initial estimate value of
$\psi_{guess}=\pi$, and repeatedly adjusts the lowerbound or upperbound to the
guess value (depending on whether $L(\psi_{guess})<\ell_{given}$ or
$\ell_{given}<L(\psi_{guess})$ respectively), until the correct angle
corresponding to $\ell_{given}$ is determined.

For the ellipse in \refeqn{3sphere-1dim-ellipse},
$d_x=\sqrt{\delta-\tfrac{\alpha^2}{3}}$ and $d_y=\tfrac{d_x}{\sqrt{3}}$.  Since
$d_x$ and $d_y$ only depend on $\delta$ and $\alpha$, the value of $\psi_1$ can
be generated and recovered (re-generated) independently of individual particle
velocities.

Note that the computational burden in sampling the ellipse occurs both
in log-based approaches and in our approach.
Thus, the forward portions incur the same computational cost.  However,
we use the same computational procedure on the reverse path,
to re-generate the sample and use it in the reverse procedure.  Since
log-based approaches rely on memory, they do not incur computational
cost on the reverse path, but incur extra memory copying cost in forward
execution.  In a normal, well-balanced parallel execution, since reversals
of collisions are far fewer than forward collisions, the extra computational
cost of our approach in the reverse procedure is much less than the
savings gained in foward execution compared to log-based approaches,
resulting in an overall reduction in run time and memory.

\subsubsection{Resolving the Geometrical Constraints}

Note that, with the sampled $\phi_1=\phi_1^{'}+\psi$,
it is possible to correctly recover the pre-collision angle $\phi_1^{'}$,
from which the values of the pre-collision velocities can be recovered,
but not necessarily their correct assignation to the identities of the
particles.  Thus, there still remains the problem of uniquely recovering the
configuration of the particles, since the preceding solution provides for two
different but equivalent configurations that only differ in that
their left and right particle identities are swapped.  Without modifying
the procedure, one additional bit of memory would be needed for each
collision to disambiguate between the two configurations.  Since the aim is to
completely eliminate memory accumulation, the model needs additional
development for reversibility, as presented next.

\refeqn{3sphere-1dim-geometry} gives the key terms in the geometrical
constraints as:
\begin{eqnarray}\label{eqn:3spheres-geometrical-terms}
\begin{aligned}
(a-b) &= \lambda\cos{\phi_1}\text{, }
(b-c) &= \lambda\cos{(\phi_1-\frac{2\pi}{3})}\text{, and }
(c-a) &= \lambda\cos{(\phi_1+\frac{2\pi}{3})}.
\end{aligned}
\end{eqnarray}

\begin{figure}[h]
  \centering
  \input{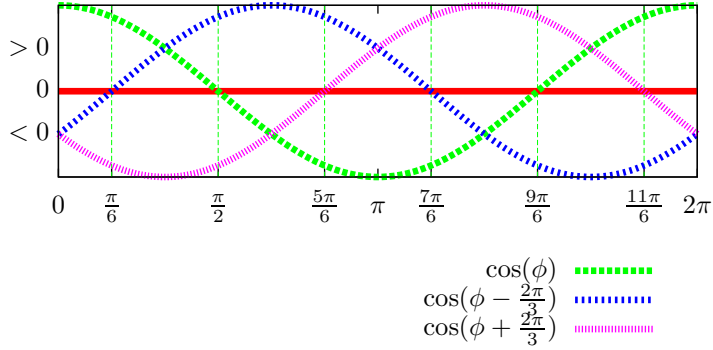}
  \caption{Variation of the geometrical terms with $\phi_1$ in 3-particle collisions in 1 dimension}
  \label{fig:3spheres-1dim-braid}
\end{figure}

\refeqn{3spheres-geometrical-terms}
indicates the permissible phase space of $\phi_1$ with the
property that we exploit here for reversibility, namely, that the three terms
$\cos\phi$, $\cos(\phi-\tfrac{2\pi}{3})$, and $\cos(\phi+\tfrac{2\pi}{3})$
never carry the same sign, i.e., if one is negative, the other two are
non-negative, and if one is positive, the other two are non-positive
(see \reffig{3spheres-1dim-braid}).

From \refeqn{3spheres-geometrical-terms}, we deduce:
\begin{eqnarray}\label{eqn:3spheres-1dim-terms}
\begin{aligned}
 \cos{\phi_1} + \cos{(\phi_1-\frac{2\pi}{3})} + \cos{(\phi_1+\frac{2\pi}{3})} = 0
 \text{, since }\lambda>0.
\end{aligned}
\end{eqnarray}

\begin{figure}[h]
\centering
\setlength{\unitlength}{0.8in}
\begin{picture}(3.25,1.0)
  \thicklines
  \put(0.5,0.5){\circle{1.0}}
  \put(0.5,0.5){\vector(1,0){0.75}}
  \put(0.55,0.40){\makebox(0,0)[t]{$a^{'}$}}
  \put(1.5,0.5){\circle{1.0}}
  \put(1.5,0.5){\vector(1,0){0.75}}
  \put(1.55,0.40){\makebox(0,0)[t]{$b^{'}$}}
  \put(2.5,0.5){\circle{1.0}}
  \put(2.5,0.5){\vector(1,0){0.75}}
  \put(2.55,0.40){\makebox(0,0)[t]{$c^{'}$}}
\end{picture}
\caption{Canonical configuration of 3 particles in 1 dimension}
\label{fig:3spheres1dimconfig}
\end{figure}
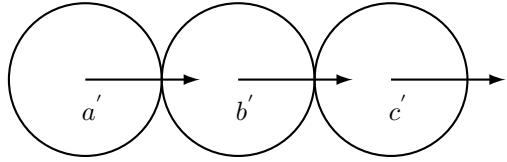

Without loss of generality, let $a^{'}$, $b^{'}$, and $c^{'}$ represent the
pre-collision velocities of the left, center and right particles,
respectively, colliding along the one-dimensional path.
This is illustrated in \reffig{3spheres1dimconfig}.   Let the
corresponding post-collision velocities be $a$, $b$, and $c$, respectively.
Define a canonical assignation of $\phi_1^{'}$ such that
$a^{'}-b^{'}=R\cos{\phi_1^{'}}$,
$b^{'}-c^{'}=R\cos{(\phi_1^{'}-\tfrac{2\pi}{3})}$, and
$c^{'}-a^{'}=R\cos{(\phi_1^{'}+\tfrac{2\pi}{3})}$.
Note that this convention fully covers
the phase space of all pre-collision velocities for the given total momentum and
kinetic energy, and in that sense, is general and equivalent to any other
valid convention.

\textbf{Pre-collision}:
For the left and center particles to collide, $a^{'}-b^{'}>0$.
Similarly, for the center and right particles to collide, $b^{'}-c^{'}>0$.
These two conditions constrain the range of
$\phi_1^{'}$ to $[\tfrac{\pi}{6},\tfrac{\pi}{2})$,
since it is only for that range of $\phi_1^{'}$ that
$\cos{\phi_1^{'}}$ and $\cos{(\phi_1^{'}-\tfrac{2\pi}{3})}$
are both non-negative.  The exact value of
$\phi_1^{'}$ in that range is determined from the following:
\begin{eqnarray}
\begin{aligned}
\bar{a}^{'} = \frac{a^{'}-b^{'}}{\sqrt{2}}\text{, }
\bar{b}^{'} = \frac{a^{'}+b^{'}}{\sqrt{2}}\text{, and }
\tan{\phi_1^{'}}=\frac{\sqrt{3}\left(\bar{b}^{'}-\frac{\sqrt{2}}{3}\alpha\right)}{\bar{a}^{'}}\text{.}
\end{aligned}
\end{eqnarray}

Thus, any pre-collision configuration of ($a^{'}$, $b^{'}$, $c^{'}$) velocities
of left, center, and right particles can be uniquely and completely represented
by ($\phi_1^{'}$, $\alpha$, $\delta$), where
$\tfrac{\pi}{6}\leq \phi_1^{'} < \tfrac{\pi}{2}$.

\textbf{Post-collision}:
For the left and center particles to diverge after
collision (i.e., not to pass through each other),
their post-collision velocities
must satisfy $a-b < 0$.  Similarly, for the center and right particles,
$b-c < 0$.  These two conditions constrain the range of $\phi_1$ to
$[\tfrac{7\pi}{6},\tfrac{3\pi}{2})$, which is essentially offset
by $+\pi$ from the range of $\phi_1^{'}$.

\begin{figure}[h]
  \centering
  \includegraphics[width=0.65\myimagewidth]{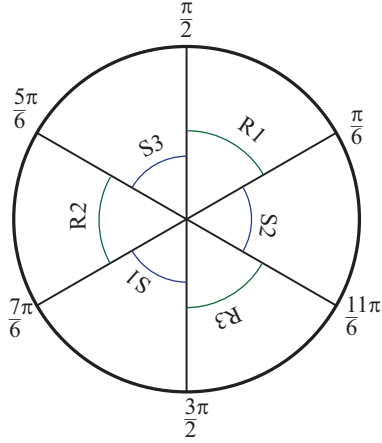}
  \caption{Division of the angular space for $\phi_1$ and $\phi_1^{'}$ in
           3-particle collisions in 1 dimension}
  \label{fig:3spheres-1dim-regions}
\end{figure}

Pictorially, the regions of interest in the $[0,2\pi)$ range are given in
\reffig{3spheres-1dim-regions}.  The full $[0,2\pi)$ range is divided into
six regions, each spanning $\tfrac{2\pi}{3}$.  The first region starts at
$\tfrac{\pi}{6}$.  In the figure, $R1$ is the range of
$\phi_1^{'}$ (pre-collision), and $S1$ is the range of $\phi_1$
(post-collision).  Any given angle $\phi_1^{'}$ in $R1$ corresponds to a set of
three velocities $\{a^{'}, b^{'}, c^{'}\}$, such that the pre-collision
geometrical constraints of \refeqn{3sphere-1dim-geometry-orig} are satisfied,
implying a specifically ordered sequence of the particles, say, $a;b;c$,
along one dimension.  The regions $R2$ and $R3$ correspond to left-rotation
of the $R1$ sequence, namely, $c;a;b$ and $b;c;a$, obtained
by offsetting $\phi_1^{'}$ by $+\tfrac{2\pi}{3}$ and $-\tfrac{2\pi}{3}$,
respectively.  Similarly, the regions $S2$ and $S3$ correspond to the
right-rotation of the $S1$ sequence.  Note that the pre-collision
and post-collision angles only fall in the $R1$ and $S1$ regions, respectively,
and the other regions are unreachable by the system.  They are defined
and used in intermediate calculations when computing post-collision
angles in forward execution (and recovering pre-collision angles in reverse)
while ensuring full phase space coverage.

\textbf{Reversible Sampling}: The problem of making the collision reversible
now becomes equivalent to defining a \textit{reversible} (i.e., one-to-one and
onto) mapping from any given $\phi_1^{'}\in [\tfrac{\pi}{6},\tfrac{\pi}{2})$ to
a random sample of $\phi_1\in [\tfrac{7\pi}{6},\tfrac{3\pi}{2})$, while
accurately preserving the underlying distribution of velocities along the
circumference of the ellipse in \refeqn{3sphere-1dim-ellipse}.  In general, any
such bijection could serve the purpose.  Here, such a mapping function is
provided in \refalgo{3part1dimforward} (forward) and \refalgo{3part1dimreverse}
(reverse).

The forward function essentially rotates $\phi_1^{'}$ uniformly over the entire
range $[0,2\pi)$ first, and then makes adjustments reversibly, as needed, to
map back to the valid range $[\tfrac{7\pi}{6},\tfrac{3\pi}{2})$ of $\phi_1$.

\begin{algorithm}[h]
    \caption{($\phi_1^{'} \longrightarrow \phi_1$): \textit{Forward Function for 3-Particle in 1 Dimension}}
    \label{algo:3part1dimforward}
    \begin{algorithmic}
        \STATE $\ell \leftarrow G\cdot L(2\pi)$ \COMMENT{Pick a random arc length on ellipse}
        \STATE $\psi_1 \leftarrow L^{-1}(\ell)$ \COMMENT{Find angle corresponding to arc length}
        \STATE \COMMENT{Compute post-collision angle}
        \STATE $\phi_1 \leftarrow (\phi_1^{'}+\psi_1) \mod 2\pi$ \COMMENT{\textbf{Step 1}}
        \STATE \COMMENT{Adjust post-collision angle if/as necessary}
        \IF {more than one of $\cos{\phi_1}$,
                              $\cos{(\phi_1-\tfrac{2\pi}{3})}$ and
                              $\cos{(\phi_1+\tfrac{2\pi}{3})}$ is positive}
            \STATE $\phi_1 \leftarrow (\phi_1+\pi) \mod 2\pi$
	    \COMMENT{\textbf{Step 2}}
        \ENDIF
        \IF {$\cos{\phi_1}$ is positive}
            \STATE $\phi_1 \leftarrow (\phi_1-\tfrac{2\pi}{3}) \mod 2\pi$
	    \COMMENT{\textbf{Step 3a}}
        \ELSIF {$\cos{(\phi_1-\tfrac{2\pi}{3})}$ is positive}
            \STATE $\phi_1 \leftarrow (\phi_1+\tfrac{2\pi}{3}) \mod 2\pi$
	    \COMMENT{\textbf{Step 3b}}
        \ENDIF
    \end{algorithmic}
\end{algorithm}

The operation of \refalgo{3part1dimforward} is illustrated in
\refeqn{3part1dimforward}.
Step 1 adds a random offset in $[0,2\pi)$ to $\phi_1^{'}$ to get a
candidate sample of $\phi_1$.  Clearly, the candidate may fall in any of the
six regions shown in \reffig{3spheres-1dim-regions}.  The first correction
is to detect if the candidate angle maps to R1, R2, or R3, and if so,
remap it to S1, S2, or S3 respectively
(because post-collision velocities must diverge, which implies that
the post-collision free angle cannot be in R1, R2, or R3).
This is accomplished in Step 2 by
adding $\pi$ to the candidate.  In Step 3, if the candidate happens to already
be in S1, then no additional adjustments are needed.  Otherwise, if it falls in
S3, it is wrapped back to S1 by rotating it counter-clockwise by
$\tfrac{2\pi}{3}$ (Step 3a), and if it falls in S2, it is wrapped back to S1 by
rotating it clockwise by $\tfrac{2\pi}{3}$ (Step 3b).  To summarize:

\begin{eqnarray}
  \phi_1^{'} \in \left\{ R1 \right\}
  \xrightarrow[\text{randomize}]{\text{Step 1}}
  \phi_1 \in
      \left\{
          S1,
          S2,
          S3,
          R1,
          R2,
          R3
      \right\}\xrightarrow[\text{maps to}]{\text{Step 2}}\cdots\notag\\
  \cdots\xrightarrow[\text{maps to}]{\text{Step 2}}
    \phi_1
    \in
      \left\{
          S1,
          S2,
          S3
      \right\}
  \xrightarrow[\text{maps to}]{\text{Step 3}}
    \phi_1 \in \left\{ S1 \right\}\text{ .}
    \label{eqn:3part1dimforward}
\end{eqnarray}

The reverse function shown in \refalgo{3part1dimreverse}
uncovers the random offset first, and reconstructs a candidate $\phi_1^{'}$.
It detects adjustments, if any, that were made during forward execution, and
then performs the opposite of adjustments to recover the correct
original value of $\phi_1^{'}$.

\begin{algorithm}[h]
    \caption{($\phi_1 \longrightarrow \phi_1^{'}$): \textit{Reverse Function for 3-Particle in 1 Dimension}}
    \label{algo:3part1dimreverse}
    \begin{algorithmic}
        \STATE $\ell \leftarrow G\cdot L(2\pi)$ \COMMENT{Recover the random arc length on ellipse}
        \STATE $\psi_1 \leftarrow L^{-1}(\ell)$ \COMMENT{Recover the angle corresponding to arc length}
        \STATE \COMMENT{Compute initial guess of pre-collision angle}
        \STATE $\phi_1^{'} \leftarrow (\phi_1-\psi_1) \mod 2\pi$
        \STATE \COMMENT{Correct the guess if/as necessary}
	\COMMENT{\textbf{Step 1}}
        \IF {more than one of $\cos{\phi_1}$,
                              $\cos{(\phi_1-\tfrac{2\pi}{3})}$ and
                              $\cos{(\phi_1+\tfrac{2\pi}{3})}$ is positive}
            \STATE $\phi_1^{'} \leftarrow (\phi_1^{'}-\pi) \mod 2\pi$
	    \COMMENT{\textbf{Step 2}}
        \ENDIF
        \IF {$\cos{\phi_1^{'}}$ is negative}
            \STATE $\phi_1^{'} \leftarrow (\phi_1^{'}-\tfrac{2\pi}{3}) \mod 2\pi$
	    \COMMENT{\textbf{Step 3a}}
        \ELSIF {$\cos{(\phi_1^{'}-\tfrac{2\pi}{3})}$ is negative}
            \STATE $\phi_1^{'} \leftarrow (\phi_1^{'}+\tfrac{2\pi}{3}) \mod 2\pi$
	    \COMMENT{\textbf{Step 3b}}
        \ENDIF
    \end{algorithmic}
\end{algorithm}

The operation of \refalgo{3part1dimreverse} is illustrated in
\refeqn{3part1dimreverse}.
Step 1 removes the random offset in $[0,2\pi)$ from $\phi_1$ to get a
first guess of the original $\phi_1^{'}$.  Clearly, the guessed value
may fall in any of the
six regions shown in \reffig{3spheres-1dim-regions}.  The first correction
to the guess is to detect if the candidate angle maps to S1, S2, or S3,
and if so, remap it to R1, R2, or R3 respectively
(because pre-collision velocities must be converging,
which implies that the pre-collision free angle
cannot be in S1, S2, or S3).
This is accomplished in Step 2 by subtracting $\pi$ from the guessed value.
In Step 3, if the guess happens to already
be in R1, then it already represent the correct original value of $\phi_1^{'}$.
Otherwise, if it falls in R2, it is wrapped back to R1 by rotating it
clockwise by $\tfrac{2\pi}{3}$
(Step 3a), and if it falls in R3, it is wrapped forward to R1 by rotating it
counter-clockwise by $\tfrac{2\pi}{3}$ (Step 3b).  To summarize:

\begin{eqnarray}
  \phi_1^{'} \in \left\{ R1 \right\}
  \xleftarrow[\text{maps to}]{\text{Step 3}}
    \phi_1^{'}
    \in
      \left\{
          R1,
          R2,
          R3
      \right\} \xleftarrow[\text{maps to}]{\text{Step 2}}\cdots\notag\\
  \cdots\xleftarrow[\text{maps to}]{\text{Step 2}}
    \phi_1^{'}
    \in
      \left\{
          S1,
          S2,
          S3,
          R1,
          R2,
          R3
      \right\}
  \xleftarrow[\text{de-randomize}]{\text{Step 1}}
  \phi_1 \in \left\{ S1 \right\}\text{ .}
  \label{eqn:3part1dimreverse}
\end{eqnarray}

\subsubsection{Combining Dynamics and Geometry}

The series of transformations performed in forward and reverse
collision operations is summarized in
\refeqn{3part1dimsummaryreverse}, which shows the input of the three
pre-collision velocities $(a',b',c')$ being transformed by the
\textbf{${\bm V}$-to-${\bm \Phi}$} mapping into the triple
formed by the angle $\phi_1^{'}$, momentum $\alpha$, and energy $\delta$.
These are fed into the forward parts of the \textit{Function 1} or
\textit{Function 2} with an additional input, which is the uniformly
distributed random number $G$.  The resulting triple of the post-collision
angle $\phi_1$ together with momentum and energy are transformed by the
\textbf{${\bm \Phi}$-to-${\bm V}$} mapping into the post-collision velocities
$(a,b,c)$.  For reversal, the forward process is inverted
by first recovering the angle $\phi_1$ upon applying the
\textbf{${\bm V}$-to-${\bm \Phi}$} mapping on the post-collision velocities.
The random number $G$ is recovered by retracing the random number sequence, and
the reverse part of the function is
applied to recover $\phi_1^{'}$, from which the pre-collision velocities
are obtained by applying the \textbf{${\bm \Phi}$-to-${\bm V}$} mapping
on the recovered angle.

\begin{eqnarray}\label{eqn:3part1dimsummaryreverse}
  \begin{aligned}
  (a^{'},b^{'},c^{'})&\xrightarrow{\textbf{${\bm V}$-to-${\bm \Phi}$}}&(\phi_1^{'},\alpha,\delta)
    &\xrightarrow{\text{Forward}}
    &(\phi_1,\alpha,\delta)&\xrightarrow{\textbf{${\bm \Phi}$-to-${\bm V}$}}&(a,b,c)\\
    & & & \quad \begin{array}{c}\rotatedleadsto{90}\\ \text{G}\\ \rotatedleadsto{270} \end{array} & &
      &{\Big \downarrow} \\
  (a^{'},b^{'},c^{'})&\xleftarrow{\textbf{${\bm \Phi}$-to-${\bm V}$}}&(\phi_1^{'},\alpha,\delta)
    &\xleftarrow{\text{Reverse}}
    &(\phi_1,\alpha,\delta)&\xleftarrow{\textbf{${\bm V}$-to-${\bm \Phi}$}}&(a,b,c)
  \end{aligned}
\end{eqnarray}

\subsection{Reversal for 3-Particle Collisions in 2 Dimensions}\label{sec:3particle-2dim}

In a collision of three particles in 2-dimensional space, the equations of
motion are:

\begin{equation}
      \left.
  \begin{aligned}
    a+b+c&=\alpha\\
    d+e+f&=\beta\\
    a^2+b^2+c^2+d^2+e^2+f^2&=\delta\\
    3\delta&>\alpha^2+\beta^2
  \end{aligned}
      \right\}
  \text{Dynamics}.
\end{equation}

The geometrical constraints satisfied by the post-collision velocities are:
\begin{equation}\label{eqn:3sphere-2dim-geometry-constraints}
      \left.
  \begin{aligned}
    r_{21x}\cdot(a-b)+r_{21y}\cdot(d-e)& > 0, &\text{if \textbf{P1} and
                                                   \textbf{P2} are in contact}\\
    r_{32x}\cdot(b-c)+r_{32y}\cdot(e-f)& > 0, &\text{if \textbf{P2} and
                                                   \textbf{P3} are in contact}\\
    r_{13x}\cdot(c-a)+r_{13y}\cdot(f-d)& > 0 &\text{if \textbf{P3} and
                                                   \textbf{P1} are in contact}\\
  \end{aligned}
      \right\}
  \text{Geometry}.
\end{equation}

The pre-collision geometrical constraints are obtained
by replacing $> 0$ by $< 0$ in the post-collision constraints.
In \refeqn{3sphere-2dim-geometry-constraints},
we will use \textbf{K1} to denote the first inequality
(\textbf{P1} and \textbf{P2} in contact),
\textbf{K2} to denote the second inequality, and \textbf{K3}
to denote the third.

The solution to the equations of motion
satisfies the hyper-ellipsoid equation:

\begin{equation}\label{eqn:3sphere-2dim-ellipsoid}
  \begin{aligned}
    {\left(\frac{\bar{a}-0}{1}\right)}^2 +
    {\left(\frac{\bar{b}-\frac{\sqrt{2}}{3}\alpha}{\dfrac{1}{\sqrt{3}}}\right)}^2 +
    {\left(\frac{\bar{d}-0}{1}\right)}^2 +
    {\left(\frac{\bar{e}-\frac{\sqrt{2}}{3}\beta}{\dfrac{1}{\sqrt{3}}}\right)}^2
    &=
      \delta-\frac{\alpha^2+\beta^2}{3}\text{,}\\
    \text{where }
    \bar{a}=\frac{a-b}{\sqrt{2}}\text{, }
    \bar{b}=\frac{a+b}{\sqrt{2}}\text{, }
    \bar{d}=\frac{d-e}{\sqrt{2}}\text{, and }
    \bar{e}=\frac{d+e}{\sqrt{2}}~.
  \end{aligned}
\end{equation}

The hyper-ellipsoid can be described via parametic equations with
three independent parameters $\left\{ \phi_1, \phi_2, \phi_3\right\}$
as the degrees of freedom as follows:

\begin{equation}\label{eqn:3sphere-2dim-ellipsoid-space}
    \left.
  \begin{aligned}
    \bar{a}&=\frac{\lambda}{\sqrt{2}}\cos{\phi_1}\text{,}\\
    \bar{d}&=\frac{\lambda}{\sqrt{2}}\sin{\phi_1}\cos{\phi_2}\text{,}\\
    \bar{b}&=\frac{\sqrt{2}}{3}\alpha+\frac{\lambda}{\sqrt{2}\sqrt{3}}\sin{\phi_1}\sin{\phi_2}\cos{\phi_3}\text{,}\\
    \bar{e}&=\frac{\sqrt{2}}{3}\beta+\frac{\lambda}{\sqrt{2}\sqrt{3}}\sin{\phi_1}\sin{\phi_2}\sin{\phi_3}\text{, where }\\
    \lambda&=\sqrt{2}\sqrt{\delta-\frac{\alpha^2+\beta^2}{3}}\text{,}
  \end{aligned}
    \right\}
  \begin{aligned}
    \phi_1&\in[0,\pi]\text{, }\\
    \phi_2&\in[0,\pi]\text{, }\\
    \phi_3&\in[0,2\pi)\text{.}\\
  \end{aligned}
\end{equation}

Based on the preceding parametric equations, the terms in the
geometrical constraints can be expressed as:

\begin{equation}\label{eqn:3sphere-2dim-diffs}
    \begin{aligned}
      a-b&=\lambda\cos{\phi_1}\\
      d-e&=\lambda\sin{\phi_1}\cos{\phi_2}\\
      b-c&=\frac{\lambda}{2}(\sqrt{3}\sin{\phi_1}\sin{\phi_2}\cos{\phi_3}-\cos{\phi_1})\\
      e-f&=\frac{\lambda}{2}(\sqrt{3}\sin{\phi_1}\sin{\phi_2}\sin{\phi_3}-\sin{\phi_1}\cos{\phi_2})\\
      c-a&=\frac{-\lambda}{2}(\sqrt{3}\sin{\phi_1}\sin{\phi_2}\cos{\phi_3}+\cos{\phi_1})\\
      f-d&=\frac{-\lambda}{2}(\sqrt{3}\sin{\phi_1}\sin{\phi_2}\sin{\phi_3}+\sin{\phi_1}\cos{\phi_2})~.\\
    \end{aligned}
\end{equation}

Note that $\sin{\phi_1}$ and $\sin{\phi_2}$ are always non-negative.

The \textbf{${\bm V}$-to-${\bm \Phi}$} and \textbf{${\bm \Phi}$-to-${\bm V}$}
mappings are obtained from \refeqn{3sphere-2dim-diffs} as follows:

\textbf{${\bm V}$-to-${\bm \Phi}$}:
Given $a-f$, the angles $\phi_1$, $\phi_2$, and $\phi_3$ are computed as:
\begin{equation}
    \begin{aligned}
      \phi_1 = & \cos^{-1}\left(\dfrac{a-b}{\lambda}\right)\\
      \phi_2 = &
          \left\{
          \begin{aligned}
            0 \text{ if $\phi_1=0$ or $\phi_1=\pi$}\\
	    \cos^{-1}\left( \dfrac{d-e}{\lambda\sin\phi_1}\right) \text{ otherwise}
          \end{aligned}
          \right.\\
      \phi_3 = &
          \left\{
          \begin{aligned}
            0 \text{ if $\phi_1=0$ or $\phi_1=\pi$ or $\phi_2=0$ or $\phi_2=\pi$}\\
	    \sin^{-1}\left( \dfrac{2(e-f)+\lambda\sin\phi_1\cos\phi_2}{\sqrt{3}\lambda\sin\phi_1\sin\phi_2}\right) \text{ otherwise}
          \end{aligned}
          \right.\\
    \end{aligned}
\end{equation}

\textbf{${\bm \Phi}$-to-${\bm V}$}:
Given $\alpha,\beta,\delta,\phi_1,\phi_2,\phi_3$,
the values of $a-f$ are computed from \refeqn{3sphere-2dim-diffs}.

Note that the cases of $\phi_1=0$ or $\phi_1=\pi$ or
$\phi_2=0$ or $\phi_2=\pi$ are solved as follows:
\begin{equation}
    \begin{aligned}
      a &= \dfrac{\alpha}{3}+\dfrac{\lambda}{2}\cos\phi_1\\
      b &= \dfrac{\alpha}{3}-\dfrac{\lambda}{2}\cos\phi_1\\
      c &= \dfrac{\alpha}{3}\\
      d &= \dfrac{\beta}{3}+\dfrac{\lambda}{2}\sin\phi_1\cos\phi_2\\
      e &= \dfrac{\beta}{3}-\dfrac{\lambda}{2}\sin\phi_1\cos\phi_2\\
      f &= \dfrac{\beta}{3}\\
    \end{aligned}
\end{equation}
Hence, when dealing with \refeqn{3sphere-2dim-ellipsoid-space}
in the remaining of the analysis, we only consider the case of $0 < \phi_1 < \pi$
and $0 < \phi_2 < \pi$.  Also, whenever $\phi_1=0$ or $\phi_1=\pi$,
we set $\phi_2=0$ and $\phi_3=0$.  Similarly, whenever $\phi_2=0$
or $\phi_2=\pi$, we set $\phi_3=0$.

\subsubsection{Possible Geometries}

To make analysis easier, a notion of a ``canonical configuration'' is
introduced for the three colliding particles in the 2-dimensional space.
The canonical view is to align the $x$-axis with the line joining the centers
of two particles in contact.  The choice of the pair chosen for this line is
designed to be recoverable in reverse execution, essentially making the choice
of the pair only dependent on the geometry of collision, independent of the
velocities of the particles undergoing the 3-particle collision.

\begin{figure}
  \centering
  \includegraphics[width=\myimagewidth,height=\myimageheight,keepaspectratio]{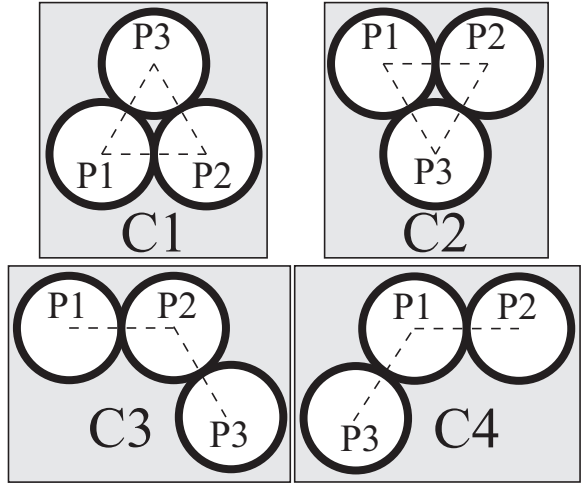}
  \caption{The canonical forms that we define for the
           four possible configurations in which three particles
           may undergo collision in 2 dimensions}
  \label{fig:3spheres-2dim-configs}
\end{figure}

When the $x$-axis is aligned along such a pair of particles in contact, they can
appear in one of four configurations shown in \reffig{3spheres-2dim-configs}.
In all configurations, the horizontal line is chosen to be the line joining
the two particles whose identifiers are smaller than that of the third one. The
particle with the smallest identifier is always chosen as the particle on the
left of the horizontal axis line.

In configurations \textbf{C1} and \textbf{C2}, all three particles are in contact
with each other, giving three pairs of particles in contact.  In each of the
rest, \textbf{C3} and \textbf{C4}, only two pairs of particles are in contact.
In all configurations, \textbf{P1} and \textbf{P2} are the left and right
particles forming the horizontal axis.  In \textbf{C1}, the third particles
\textbf{P3} is above the two horizontally placed particles, while, in
\textbf{C2}, the third particle is below them.  The configurations \textbf{C3}
and \textbf{C4} cover the rest of the possibilities in which only two pairs of
the particles are in contact with each other at the same time.  In \textbf{C3},
the third particle \textbf{P3} is only in contact with particle \textbf{P2}, while,
in \textbf{C4}, \textbf{P3} is only in contact with particle \textbf{P1}.

Once the canonical configuration is chosen, all the velocities are rotated to
reorient to the new axes.  Total momenta and energy undergo a resultant change
that can be reversed to recover original momenta and energy values simply by
rotating the axes back to original axes.  Since the original velocities
are in a one-to-one relation with the transformed velocities, it is the
transformed
velocities that will be considered as velocities $a..f$ defined earlier for
the 3-particle, 2-dimension collision model.  After computing
individual post-collision velocities, they are rotated back to the original
frame of reference, thus restoring the original total momenta and energy.

\subsubsection{Sampling the ${\bm \Psi}$ Phase Space}

To generate a random set of offsets, ${\bm \Psi}$, a random sample point
on the hyper-ellipsoid of
\refeqn{3sphere-2dim-ellipsoid} is generated by invoking the numerical
method given in \refalgo{hyperellipsoidsampling} with $s=4$, and
$\{{}_4\lambda_i\mid 1\leq i\leq 4\}$ obtained by converting
\refeqn{3sphere-2dim-ellipsoid} into the canonical form expected by the
algorithm.  The sampling approach is a generalized version of
the approach employed for 3-particle collisions in 1 dimension
(\refsec{3spheres-1dim}) to sample the phase space of ${\bm \Psi}$.  As
mentioned earlier, although rejection-based methods are available for
generating the samples, rejection is incompatible with reversibility, making
them inapplicable here.

\begin{algorithm}[h]
 \caption{(${\bm G} \longrightarrow {\bm \Psi}$): \textit{Generate the Parameters ${\bm \Psi}$ of a Random Point on the Surface of an $s$-Dimensional Hyper-Ellipsoid, ${\cal H}_s$, using Random Numbers ${\bm G}=\{G_1,\ldots,G_{s-1}\}$}}
    \label{algo:hyperellipsoidsampling}
    \begin{algorithmic}[1]
        \STATE \textbf{Input}: $s, \{{}_s\lambda_i \mid 1 \leq i \leq s\}$, where integer $s>1$, and $\sum_{i=1}^s\left(\tfrac{{}x_i}{{}_s\lambda_i}\right)^2=1$ is the hyper-ellipsoid
        \STATE \textbf{Output}: $\{\psi_{i} \mid 1 \leq i < s\}$, where $\psi_i$ are the parameters of a random point $({}_rx_1,\ldots,{}_rx_s)$ on the hyper-ellipsoid, such that ${}_rx_i={}_s\lambda_i\cos{\psi_i}\prod_{j=1}^{i-1}{\sin{\psi_j}} \text{ for all } 1\leq i<s$, and ${}_rx_s={}_s\lambda_s\prod_{j=1}^{s}{\sin{\psi_j}}$
        \FOR{$k=s$ \textbf{down-to} $2$}
            \STATE Let $\overline{k} = s-k+1$
            \IF {$k=2$}
                \STATE Invoke the algorithm in \refsec{ellipsecircumf}
                       using $G_{\overline{k}}$
		       to generate the parameter $0\leq\psi^*<2\pi$
                       corresponding to a random point on the perimeter of
                       the (2-dimensional) ellipse ${\cal H}_2$
                \STATE $\psi_{\overline{k}} \leftarrow \psi^*$
            \ELSE
                \STATE Compute the surface area $A_k$ of ${\cal H}_k$
                \STATE Compute the random fraction ${}_rA_k$ of $A_k$ as:
                       ${}_rA_k \leftarrow G_{\overline{k}}\cdot A_k$
                \STATE Using the bisection method analogous to that in
                       \refsec{ellipsecircumf}, determine a
		       $0\leq\psi^*\leq\pi$ such that surface area of
		       the calotte of ${\cal H}_k$ defined by the
		       latitudinal angle $\psi^*$ from
		       the pole equals ${}_rA_k$
                \STATE $\psi_{\overline{k}} \leftarrow \psi^*$
                \STATE Determine the parameters
		       $\{{}_{s-1}\lambda_i\mid 1\leq i\leq s-1\}$
		       of the $(s-1)$-dimensional ellipsoid ${\cal H}_{s-1}$
		       formed by the opening of the calotte, obtained
		       by substituting
		       ${}_sx_{\overline{k}}={}_s\lambda_{\overline{k}}\cos\psi_{\overline{k}}$.
            \ENDIF
        \ENDFOR
    \end{algorithmic}
\end{algorithm}

Note that, by using one extra random number (i.e., using $d_n+1$ random
numbers instead of $d_n$),
\refalgo{hyperellipsoidsampling} can be elegantly generalized, avoiding
reliance on the separate algorithm of \refsec{ellipsecircumf} for the special
case of a 2-dimensional ellipsoid (ellipse).  This can be achieved by iterating
down to $s=1$, introducing an extra parameter $\psi_s$, and using the
additional random number to set $\psi_s$ to either $\tfrac{\pi}{2}$ or
$\tfrac{3\pi}{2}$ (\textit{i.e.}, ${}_1x_1=\pm{}_1\lambda_1$).
Using such a generalized algorithm
for sampling the surface of an $s$-dimensional ellipsoid, $s\geq 1$,
the algorithm in
\refsec{ellipsecircumf} can be replaced by a call to the generalized
\refalgo{hyperellipsoidsampling} with $s=2$.  However, to restrict the number
of random numbers to $d_n$, we customize the last iteration to use the
optimized version given in \refsec{ellipsecircumf}.

In the next sections, we derive the permissible ranges of the parameters,
restricted from their full, nominal ranges due to conservation laws as well
as geometrical constraints, and develop the forward and reverse collision
procedures based on the derived parameter ranges.

\subsubsection{Configuration C1}

\begin{figure}[h]
  \centering
  \includegraphics[width=0.5\myimagewidth,height=0.5\myimageheight,keepaspectratio]{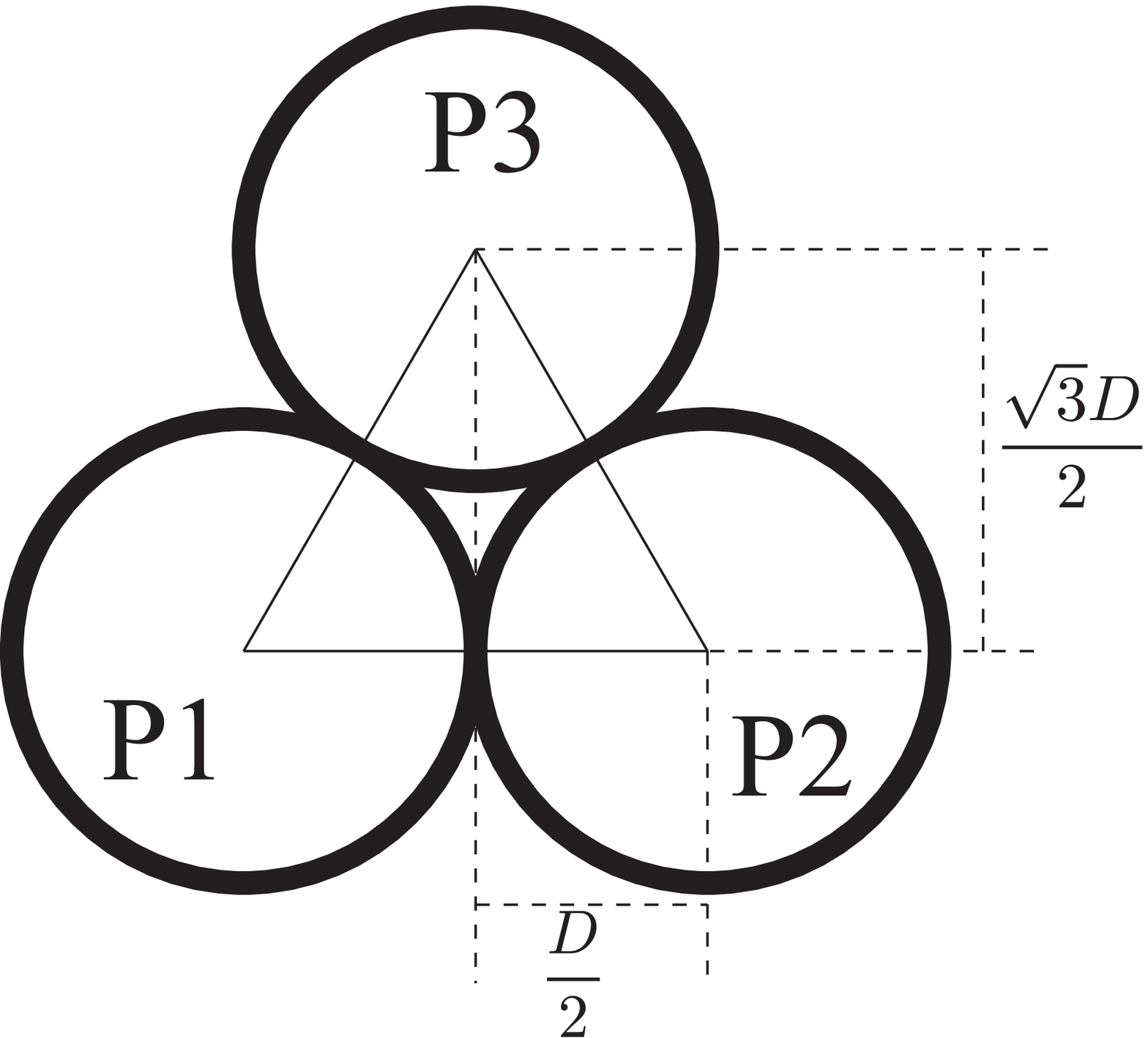}
  \caption{Measures of interest in configuration \textbf{C1}}
  \label{fig:3spheres-2dim-config1}
\end{figure}

From the geometry of \textbf{C1} (ref. \reffig{3spheres-2dim-config1}),
it can be seen that $r_{21x}=D$,
$r_{21y}=0$, $r_{32x}=\tfrac{-D}{2}$, $r_{32y}=\tfrac{\sqrt{3}}{2}D$,
$r_{13x}=\tfrac{-D}{2}$, and $r_{13y}=\tfrac{-\sqrt{3}}{2}D$.
Using these in \refeqn{3sphere-2dim-geometry-constraints} and
\refeqn{3sphere-2dim-diffs} to account for the geometrical
constraints, we obtain the ranges of $\phi_1$ and $\phi_2$
that are more constrained than in \refeqn{3sphere-2dim-ellipsoid-space}.

The inequality \textbf{K1} directly gives the following:
\begin{equation}\label{eqn:3spheres-2dim-c1-deriv1}
  \begin{aligned}
    D\cdot(\lambda\cos{\phi_1})+0\cdot(\cdots) > 0\\
    \implies \cos{\phi_1}\geq 0\\
    \implies -\tfrac{\pi}{2} \leq \phi_1 \leq \tfrac{\pi}{2}\\
    \implies 0 \leq \phi_1 < \tfrac{\pi}{2} \text{ from \refeqn{3sphere-2dim-ellipsoid-space}.}\\
  \end{aligned}
\end{equation}

Inequality \textbf{K2} gives:
\begin{equation}\label{eqn:3spheres-2dim-c1-deriv2-1}
  \begin{aligned}
    \tfrac{-D}{2}\cdot\left(\tfrac{\lambda}{{2}}(\sqrt{3}\sin{\phi_1}\sin{\phi_2}\cos{\phi_3}-\cos{\phi_1})\right)+\\
    \tfrac{\sqrt{3}D}{2}\cdot\left(\tfrac{\lambda}{{2}}(\sqrt{3}\sin{\phi_1}\sin{\phi_2}\sin{\phi_3}-\sin{\phi_1}\cos{\phi_2})\right) > 0.\\
    \text{or, }
    -\sqrt{3}\sin{\phi_1}\sin{\phi_2}\cos{\phi_3}+\cos{\phi_1}+
    3\sin{\phi_1}\sin{\phi_2}\sin{\phi_3}-\sqrt{3}\sin{\phi_1}\cos{\phi_2} > 0.\\
  \end{aligned}
\end{equation}

Inequality \textbf{K3} gives:
\begin{equation}\label{eqn:3spheres-2dim-c1-deriv2-2}
  \begin{aligned}
    \tfrac{-D}{2}\cdot\left(\tfrac{-\lambda}{{2}}(\sqrt{3}\sin{\phi_1}\sin{\phi_2}\cos{\phi_3}+\cos{\phi_1})\right)+\\
    \tfrac{-\sqrt{3}D}{2}\cdot\left(\tfrac{-\lambda}{{2}}(\sqrt{3}\sin{\phi_1}\sin{\phi_2}\sin{\phi_3}+\sin{\phi_1}\cos{\phi_2})\right) > 0.\\
    \text{or, }
    \sqrt{3}\sin{\phi_1}\sin{\phi_2}\cos{\phi_3}+\cos{\phi_1}+
    3\sin{\phi_1}\sin{\phi_2}\sin{\phi_3}+\sqrt{3}\sin{\phi_1}\cos{\phi_2} > 0.\\
  \end{aligned}
\end{equation}

\remove{ 
From \textbf{K2}, we get:
\begin{equation}\label{eqn:3spheres-2dim-c1-l1}
  \begin{aligned}
    \text{\textbf{L1}:~ ~ ~} & \cos\phi_1 + 3\sin\phi_1\sin\phi_2\sin\phi_3 > \sqrt{3}\sin\phi_1(\sin\phi_2\cos\phi_3+\cos\phi_2)\\
    \implies & \tfrac{1}{\sqrt(3)\tan\phi_1\sin\phi_2}-\tfrac{1}{\tan\phi_2} > \cos\phi_3-\sqrt{3}\sin\phi_3\\
    \implies & \cos(\phi_3+\tfrac{\pi}{3}) < \tfrac{1}{2\sqrt{3}\tan\phi_1\sin\phi_2}-\tfrac{1}{2\tan\phi_2}.
  \end{aligned}
\end{equation}
}

From \textbf{K2}, we get:
\begin{equation}\label{eqn:3spheres-2dim-c1-l1}
  \begin{aligned}
    \text{\textbf{L1}:~ ~ ~} & \cos(\phi_3+\tfrac{\pi}{3}) < \tfrac{1}{2\sqrt{3}\tan\phi_1\sin\phi_2}-\tfrac{1}{2\tan\phi_2}.
  \end{aligned}
\end{equation}
Similarly, from \textbf{K3}, we get:
\begin{equation}\label{eqn:3spheres-2dim-c1-l2}
  \begin{aligned}
    \text{\textbf{L2}:~ ~ ~} & -\cos(\phi_3-\tfrac{\pi}{3}) < \tfrac{1}{2\sqrt{3}\tan\phi_1\sin\phi_2}+\tfrac{1}{2\tan\phi_2}.
  \end{aligned}
\end{equation}

To ensure a valid range for the left hand side in \textbf{L1},
the right hand side (RHS) of the same must not be less than $-1$.  Setting
the RHS to $-1$ defines the boundary between the possible and
impossible regions, in terms of the relation between $\phi_1$ and $\phi_2$.
Similar restrictions arise from \textbf{L2}.
These considerations give the limits on $\phi_1$ and $\phi_2$ as follows:
\begin{equation}\label{eqn:3spheres-2dim-c1-rhs-1}
  \begin{aligned}
    \phi_1 & = & \cot^{-1}(\sqrt{3}\cos\phi_2-2\sqrt{3}\sin\phi_2) & \text{ (from \textbf{L1}, RHS=$-1$)}\\
    \phi_1 & = & \cot^{-1}(-\sqrt{3}\cos\phi_2-2\sqrt{3}\sin\phi_2) & \text{ (from \textbf{L2}, RHS=$-1$)}.
  \end{aligned}
\end{equation}
When $0\leq \phi_1\leq \phi_1^* = \tfrac{\pi}{6}$,
$\phi_2$ is unrestricted in its range of $[0,\pi]$.
When $\phi_1 > \phi_1^* = \tfrac{\pi}{6}$,
the lower- and upper bounds of $\phi_2$ are restricted, as determined next.
Let $r=\tfrac{\cot\phi_1}{\sqrt{3}}$ (giving $r<1$ when
$\tfrac{\pi}{6}<\phi_1<\tfrac{\pi}{2}$).
Then, the lower bound $\phi_{2_l} \leq \phi_2$ is obtained by solving
$r=\cos\phi_2-2\sin\phi_2$, giving
$\phi_{2_l} = 2\tan^{-1}(\tfrac{-2+\sqrt{5-r^2}}{1+r})$.
Similarly, the upper bound $\phi_2 \leq \phi_{2_u}$ is obtained by solving
$r=-\cos\phi_2-2\sin\phi_2$, giving
$\phi_{2_u} = 2\tan^{-1}(\tfrac{+2+\sqrt{5-r^2}}{1-r})$.

Also, the values of $\phi_1$ and $\phi_2$ could restrict the range of $\phi_3$,
whose limits are obtained by setting the RHS to unity.
The restrictions on $\phi_3$ are obtained from:
\begin{equation}\label{eqn:3spheres-2dim-c1-rhs1}
  \begin{aligned}
    \phi_1 & = & \cot^{-1}(\sqrt{3}\cos\phi_2+2\sqrt{3}\sin\phi_2) & \text{ (from \textbf{L1}, RHS=1)}\\
    \phi_1 & = & \cot^{-1}(-\sqrt{3}\cos\phi_2+2\sqrt{3}\sin\phi_2) & \text{ (from \textbf{L2}, RHS=1)}.
  \end{aligned}
\end{equation}

\begin{figure}[h]
  \centering
  \input{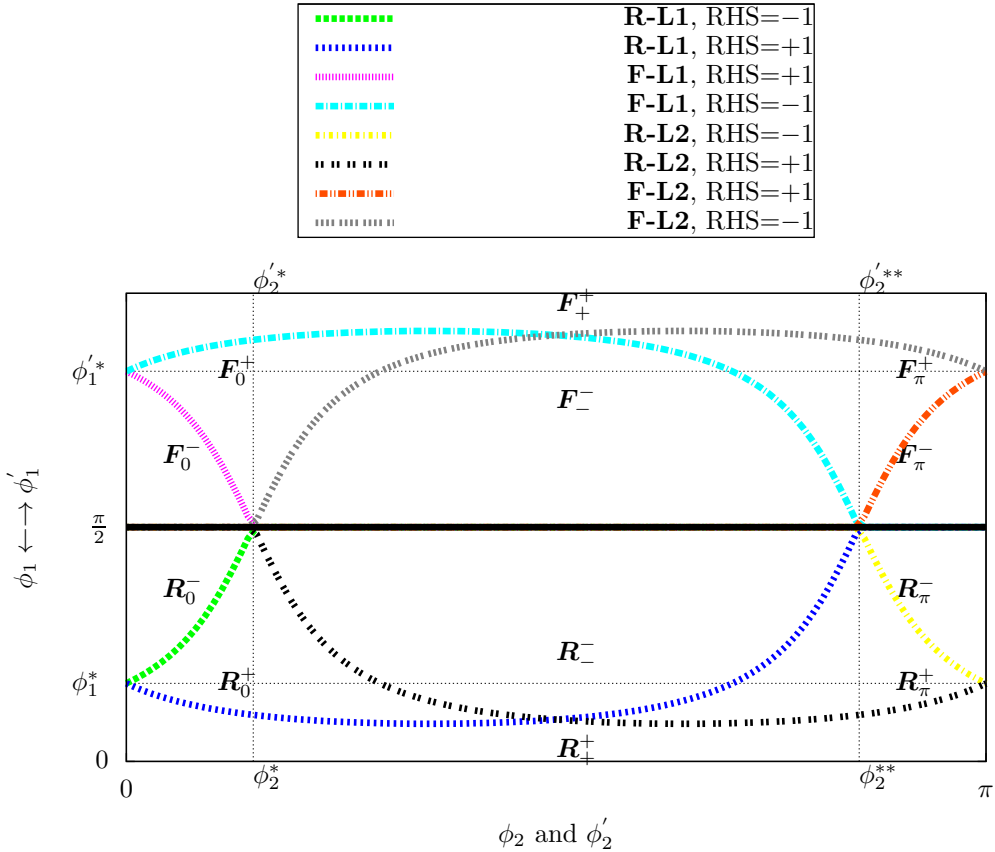}
  \caption{Regions in the nominal phase spaces of $\phi_1$ and $\phi_2$ demarcating different bounds of $\phi_1$, $\phi_2$, and $\phi_3$ in Configuration 1 of 3-particle collisions in 2 dimensions}
  \label{fig:3spheres-2dim-c1-phiranges}
\end{figure}

All the limiting curves on the angles are
illustrated in \reffig{3spheres-2dim-c1-phiranges}, which shows
the space spanned by the nominal ranges of $\phi_1\in [0,\tfrac{\pi}{2}]$
and $\phi_2\in [0,\pi]$.  The space is divided into six different
regions that impose different constraints on the ranges of $\phi_1$, $\phi_2$
and $\phi_3$.  The first two regions, labeled ${\bm R^-_0}$ and ${\bm R^-_{\pi}}$, demarcate
the regions excluded to make RHS$>-1$ (\refeqn{3spheres-2dim-c1-rhs-1}),
one on each side of $\phi_2=0$ and $\phi_2=\pi$.  In the region marked ${\bm R^+_0}$,
the range of $\phi_3$ is restricted from below to be greater than 0,
and in the region marked ${\bm R^+_{\pi}}$, the range of $\phi_3$ is restricted from
the above to be less than $2\pi$.  In the region marked ${\bm R^-_-}$, $\phi_3$ is
restricted from both below and above.  The appropriate lower bound $\phi_{3_l}$,
and upper bound $\phi_{3_u}$ on $\phi_3$ can be computed accordingly.
 In the region marked ${\bm R^+_+}$, $\phi_3$'s original range of $[0,2\pi)$ is
unrestricted.

Thus, for \textbf{C1}, the ranges for post-collision parameters are:
\begin{equation}\label{eqn:3spheres-2dim-c1-ranges}
  \begin{aligned}
   \phi_1 & \in
       [0, \tfrac{\pi}{2})\\
   \phi_2 & \in
       \left\{
       \begin{aligned}[]
	 [0, 0] & \text{ if }\phi_1 = 0\\
	 [0, \pi] & \text{ if } 0 < \phi_1 \leq \tfrac{\pi}{6}\\
	 [\phi_{2_l}, \phi_{2_u}] & \text{ otherwise } (\textit{i.e.}, \tfrac{\pi}{6}<\phi_1 < \tfrac{\pi}{2})
       \end{aligned}
       \right.\\
   \phi_3 & \in
       \left\{
       \begin{aligned}[]
	 [0, 0] & \text{ if }\phi_1 = 0 \text{ or } \phi_2 = 0 \text{ or } \phi_2 = \pi\\
	 [\phi_{3_l}, \phi_{3_u}] & \text{ if ($\phi_1$,$\phi_2$) falls in ${\bm R^+_0}$, ${\bm R^+_\pi}$ or ${\bm R^-_-}$}\\
	 [0, 2\pi) & \text{ otherwise (\textit{i.e.}, ($\phi_1$,$\phi_2$) falls in ${\bm R^+_+}$).}
       \end{aligned}
       \right.\\
  \end{aligned}
\end{equation}

Using a similar analysis, the pre-collision ranges can be obtained, as
illustrated in \reffig{3spheres-2dim-c1-phiranges}, which shows
the range of $\phi_1^{'} \in (\tfrac{\pi}{2},\pi]$, and corresponding
constraints of $\phi_2^{'}$ and $\phi_2^{'}$ in terms of the \textit{forward}
regions ${\bm F_0^-}$, ${\bm F_\pi^-}$, and so on.

The forward and reverse procedures for this configuration are given in
\refalgo{3part2dimc1forward} and \refalgo{3part2dimc1reverse} respectively.

\begin{algorithm}[h]
 \caption{($(\phi_1^{'}, \phi_2^{'}, \phi_3^{'}) \longrightarrow (\phi_1,\phi_2,\phi_3)$): \textit{Forward Procedure for Configuration 1 of a 3-Particle Collision in 2 dimensions}}
    \label{algo:3part2dimc1forward}
    \begin{algorithmic}[1]
        \STATE Invoke \refalgo{hyperellipsoidsampling} to generate the
               parameters $(\psi_1, \psi_2, \psi_3)$ of a random point
               on surface of the hyper-ellipsoid represented by
               \refeqn{3sphere-2dim-ellipsoid}
        \STATE $\phi_1\leftarrow (\phi_1^{'}-\tfrac{\pi}{2}+\psi_1) \mod \tfrac{\pi}{2}$
        \STATE $\phi_2\leftarrow (\phi_2^{'}+\psi_2) \mod \pi$
        \STATE Compute $\phi_{2_l}$ and $\phi_{2_u}$ \COMMENT{based on $\phi_1$}
	\IF {$\phi_2 < \phi_{2_l}$ \OR $\phi_{2_u} < \phi_2$}
            \STATE $\phi_2 \leftarrow (\phi_2 + (\phi_{2_l} + \pi - \phi_{2_u})) \mod \pi$
        \ENDIF
        \STATE Compute $\phi_{3_l}$ and $\phi_{3_u}$ \COMMENT{based on $\phi_1$ and $\phi_2$}
        \STATE $\phi_3\leftarrow(\phi_3^{'}+\psi_3) \mod 2\pi$
	\IF {$\phi_3 < \phi_{3_l}$}
            \STATE $\phi_3 \leftarrow (\phi_3 + \phi_{3_l}) \mod 2\pi$
	\ELSIF {$\phi_3 > \phi_{3_u}$}
            \STATE $\phi_3 \leftarrow (\phi_3 + (2\pi-\phi_{3_u})) \mod 2\pi$
        \ENDIF
    \end{algorithmic}
\end{algorithm}

\begin{algorithm}[h]
 \caption{($(\phi_1,\phi_2,\phi_3) \longrightarrow (\phi_1^{'}, \phi_2^{'}, \phi_3^{'})$): \textit{Reverse Procedure for Configuration 1 of a 3-Particle Collision in 2 dimensions}}
    \label{algo:3part2dimc1reverse}
    \begin{algorithmic}[1]
        \STATE Invoke \refalgo{hyperellipsoidsampling} to \textit{re}-generate
               the parameters $(\psi_1, \psi_2, \psi_3)$ of the random point
               on surface of the hyper-ellipsoid represented by
               \refeqn{3sphere-2dim-ellipsoid},
               previously generated by \refalgo{3part2dimc1forward}
        \STATE Recompute $\phi_{2_l}$ and $\phi_{2_u}$ \COMMENT{based on $\phi_1$}
        \STATE Recompute $\phi_{3_l}$ and $\phi_{3_u}$ \COMMENT{based on $\phi_1$ and $\phi_2$}
        \STATE $\phi_1^{'}\leftarrow \tfrac{\pi}{2}+((\phi_1-\psi_1) \mod \tfrac{\pi}{2})$
        \STATE $\phi_2^{'}\leftarrow (\phi_2-\psi_2) \mod \pi$
	\IF {$\phi_2^{'} < \phi_{2_l}$ \OR $\phi_{2_u} < \phi_2^{'}$}
            \STATE $\phi_2^{'} \leftarrow (\phi_2^{'} - (\phi_{2_l} + \pi - \phi_{2_u})) \mod \pi$
        \ENDIF
        \STATE $\phi_3^{'}\leftarrow(\phi_3-\psi_3) \mod 2\pi$
	\IF {$\phi_3^{'} < \phi_{3_l}$}
            \STATE $\phi_3^{'} \leftarrow (\phi_3^{'} - \phi_{3_l}) \mod 2\pi$
	\ELSIF {$\phi_3^{'} > \phi_{3_u}$}
            \STATE $\phi_3^{'} \leftarrow (\phi_3^{'} - (2\pi-\phi_{3_u})) \mod 2\pi$
        \ENDIF
    \end{algorithmic}
\end{algorithm}

\subsubsection{Configuration C2}

Using algebra similar to that
for \textbf{C1}, the ranges for configuration \textbf{C2} are obtained,
with swapped signs for $r_{32y}$ and $r_{13y}$.

\subsubsection{Configuration C3}

\begin{figure}[h]
  \centering
  \includegraphics[width=0.5\myimagewidth,height=0.5\myimageheight,keepaspectratio]{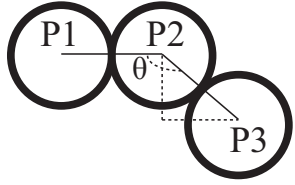}
  \caption{Parametrization of configuration \textbf{C3} by angle $\theta$, $\tfrac{\pi}{3}<\theta<\tfrac{5\pi}{3}$}
  \label{fig:3spheres-2dim-config3}
\end{figure}

Configuration \textbf{C3} can be parameterized by an angle
$\theta$ that \textbf{P3} makes relative to \textbf{P2}, as shown in
\reffig{3spheres-2dim-config3}.

In this configuration, $r_{21x}=D$, $r_{21y}=0$ (as was the case for
\textbf{C1} and \textbf{C2}), but $r_{32x}=-D\cos{\theta}$,
$r_{32y}=-D\sin{\theta}$, and we are not concerned about $r_{13x}$ and
$r_{13y}$.  Using these in \refeqn{3sphere-2dim-geometry-constraints}, we get
the geometrically constrained ranges of $\phi_1$, $\phi_2$, and $\phi_3$.
Since the inquality \textbf{K1} for this configuration is the same
as for configurations \textbf{C1} and \textbf{C2}, the range of $\phi_1$
remains $(0,\tfrac{\pi}{2}]$.
However, only the inequality \textbf{K2}
applies to this configuration, and \textbf{K3} need not apply.
From \textbf{K2}, we get:
\begin{equation}\label{eqn:3spheres-2dim-c3-deriv2}
  \begin{aligned}
  \nonumber
  -D\cos\theta\tfrac{\lambda}{{2}}(\sqrt{3}\sin\phi_1\sin\phi_2\cos\phi_3-\cos\phi_1)\\
  -D\sin\theta\tfrac{\lambda}{{2}}(\sqrt{3}\sin\phi_1\sin\phi_2\sin\phi_3-\sin\phi_1\cos\phi_2) > 0 \text{ .}\\
  \end{aligned}
\end{equation}
Since $\sin\phi_1\neq0$ and $\sin\phi_2\neq0$,
\begin{equation}\label{eqn:3spheres-2dim-c3-deriv3}
  \begin{aligned}
   & \cos(\phi_3-\theta)\sqrt{3}\sin\phi_1\sin\phi_2 < \cos\theta\cos\phi_1+\sin\theta\sin\phi_1\cos\phi_2\\
   \implies & \cos(\phi_3-\theta) < \gamma \text{, where }\\
   & \gamma=\dfrac{\cos\theta\cos\phi_1 + \sin\theta\sin\phi_1\cos\phi_2}{\sqrt{3}\sin\phi_1\sin\phi_2}\\
    \implies & \phi_3 \in [\phi_{3_l}+\theta, \phi_{3_u}+\theta] \text{, where}\\
    & 0 \leq (\phi_{3_l}+\theta)\bmod{2\pi} \leq
      (\phi_{3_u}+\theta)\bmod{2\pi} \leq 2\pi \text{, and }\\
  & \phi_{3_l} \text{ and } \phi_{3_u} \text{ are solutions to }\cos^{-1}\gamma.\\
  \end{aligned}
\end{equation}
Also, the range of $\phi_2$ may be constrained due to the requirement
that $-1 \leq \gamma \leq 1$.
Let $\mu=\tfrac{\cos\theta}{\sqrt{3}\tan\phi_1}$ and
$\nu=\tfrac{\sin\theta}{\sqrt{3}}$.
Then, $\gamma=\tfrac{\mu+\nu\cos\phi_2}{\sin\phi_2}$.
If $\mu-\nu\geq 0$ or $\mu+\nu\leq 0$, then
the range of $\phi_2$ is not constrained, giving the lowerbound
$\phi_{2_l}\leq \phi_2$ equal to 0 and upperbound $\phi_2\leq\phi_{2_u}$
equal to $\pi$.  Otherwise, the lower bound
(greater than $0$) and upper bound (less than $\pi$) of $\phi_2$ must be
determined as follows.

\begin{figure}[h]
  \centering
  \input{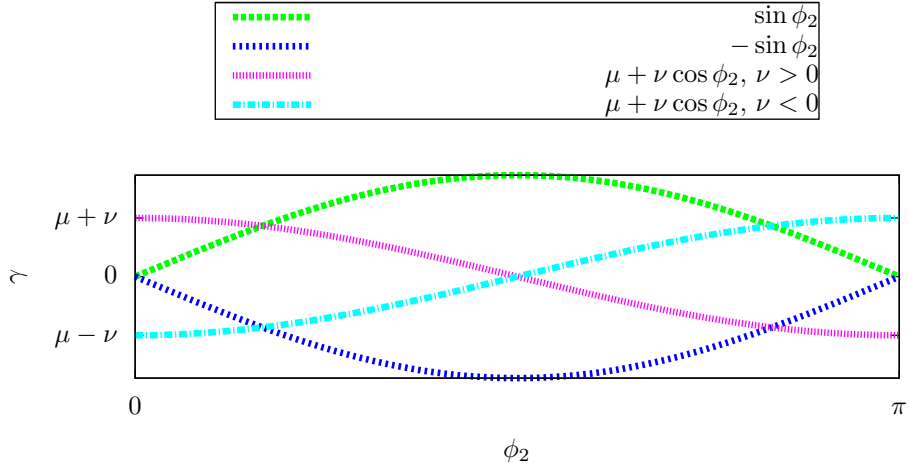}
  \caption{Illustration of lower and upper bounds of $\phi_2$ in configuration 3 of
           3-particle collisions in 2 dimensions}
  \label{fig:3spheres-2dim-c3-ineq1}
\end{figure}

If $\mu-\nu<0$, the lowerbound of $\phi_2$ is obtained by
solving for $\phi_2$ in $-\sin\phi_2=\mu+\nu\cos\phi_2$.
Since $\sin\phi_2$ and $\mu+\nu\cos\phi_2$ intersect
in at most two points within the range $(0,\pi)$ (see \reffig{3spheres-2dim-c3-ineq1}),
a unique lowerbound $\phi_{2_l}$,
$0 < \phi_{2_l} \leq \phi_2 < \pi$ is obtainable.
Similarly, if $\mu+\nu>0$, then a unique upperbound $\phi_{2_u}$,
$0 < \phi_2 \leq \phi_{2_u} < \pi$ is obtained by solving
for $\phi_2$ in $\sin\phi_2=\mu+\nu\cos\phi_2$.

Thus, for \textbf{C3}, the ranges are:
\begin{equation}\label{eqn:3spheres-2dim-c3-ranges}
  \begin{aligned}[]
   \phi_1 & \in
       [0, \tfrac{\pi}{2})\\
   \phi_2 & \in
       \left\{
       \begin{aligned}[]
	 [0, 0] & \text{ if }\phi_1 = 0\\
	 [0, \pi] & \text{ if } \mu-\nu\geq 0 \text{ or } \mu+\nu\leq 0\\
	 [\phi_{2_l}, \phi_{2_u}] & \text{ otherwise }
       \end{aligned}
       \right.\\
   \phi_3 & \in
       \left\{
       \begin{aligned}[]
	 [0, 0] & \text{ if }\phi_1 = 0 \text{ or } \phi_2 = 0 \text{ or } \phi_2 = \pi\\
	 [\phi_{3_l}+\theta, \phi_{3_u}+\theta] & \text{ otherwise.}
       \end{aligned}
       \right.\\
  \end{aligned}
\end{equation}

The ranges may be derived similarly for pre-collision.
The range of $\phi_1^{'}$ remains
to be the same as in configuration 1 at $(\tfrac{\pi}{2},\pi]$,
but, \refeqn{3spheres-2dim-c3-deriv3} changes to
$\cos{(\phi_3^{'}-\theta)} > \gamma$.

\subsubsection{Configuration C4}

The treatment of configuration \textbf{C4} proceeds similar to that
for configuration \textbf{C3}, using \textbf{K3} instead of \textbf{K2}.

\subsection{Random Number Generation}\label{sec:rngstreams}

A subtle but important consideration in zero-memory reversal is
the need to ensure that the exact number of random number invocations
is also recovered during reverse execution without explicitly storing that
information.  In other words, the
number of random numbers thrown, $G_C$, for any collision $C$ must be
determinable by the collision operator such that the
random number sequence is correctly reinstated to the proper position
corresponding to the pre-collision state.  All our algorithms possess
this property, with $G_C=d_n$.

In the case of 2-particle collisions in 1 dimension, no
random numbers are needed, trivially satisfying the property.  In the
case of 2-particle collisions in 2 dimensions, exactly one random number is
generated per collision (\refalgo{2part2dimforward}),
and hence the random number stream is stepped
back by exactly one number during reversal of that collision
(\refalgo{2part2dimreverse}).

The case of 2-particle collisions in
3 dimensions is slightly more complex, since it requires more than one
random number to be generated.  For forward collision,
it appears possible to sometimes generate one random number and some other times
two.  Only one random number (let us denote it by $G_1$) appears
sufficient to be generated if that number happens to result in
$\phi_1=0$ or $\phi_1=\pi$.  Similarly, two random numbers (let us denote
them by $G_1$ and $G_2$) appear needed only
to generate $\phi_2$ if the randomly generated $\phi_1$ (from $G_1$)
is such that $\phi_1\neq 0$ and $\phi_1\neq\pi$.  However, this conditional
generation of one or two random numbers per collision creates difficulties
during reversal, because, when the random number stream is reversed and
the previous random number $G$ is recovered, we will remain unsure whether
$G$ corresponds to $G_1$ or $G_2$.  It is impossible to disambiguate between
the two possibilities because both $\phi_1$ and $\phi_2$ may assume the value
of zero.  Hence, we would remain unsure if, in the forward collision, $G_1$
was zero and hence $G_2$ was not used for that collision, or if $G_1$
happened to be non-zero and $G_2=G$ happened to be zero.  Similar ambiguity
can be argued for the case of $G\neq 0$.  Due to these considerations, we
fix the number of random numbers generated per forward collision
to be exactly two, unconditionally, so that the random number stream
can be reversed exactly by two, restoring it to the correct pre-collision
state.  If $G_1$ results in $\phi_1=0$, $G_2$ is still generated from the
stream, but simply discarded by the collision algorithm.
Assuming that the stream is random, the discarding of $G_2$
when $\phi_1=0$ does not affect the uniformity of the random samples.
Note that the discarding is performed \textit{unconditionally}, without any
dependence or usage of the actual value of the discarded $G_2$ in the forward
model.  This aspect of unconditional discarding is crucial for reversal,
because the reversal also can determinstically reverse the random number
stream.

For 3-particle collisions in 1 dimension, exactly one random angle is required
for every forward collision, and hence one reversal is necessary and sufficient
in each reverse collision, ensuring correct reversal of the random stream.
For 3-particle collisions in 2 dimensions (\refappsec{3particle-2dim}),
one, two, or three random angles are needed (depending on the geometry
and pre-collision velocities) per forward collision.  However, due to
considerations similar to those for
the case of 2-particle collisions in 3 dimensions, we use exactly three
random numbers for every forward collision (even if only one or two
of them may be sufficient in special cases of dynamics and geometry)
in order to reverse the random number stream correctly.

In general, exactly $d_n$ random numbers must be generated for every collision
involving $n$ particles in $d$ dimensions.

With regard to the number of distinct streams to employ in the simulation,
it is possible to use one of the following three approaches:
(1) a single random number stream for the entire system, or
(2) $N$ independent streams, corresponding to each particle in the system, or
(3) $d_n$ independent streams for use in each collision.  The first
approach clearly requires a very high quality random number generator with
a very long period in order to support a large number of collisions when
$N$ is large.  The second approach requires relatively smaller periods per
stream but also requires minimal correlation between streams.
In any given collision, the random stream of the particle with
the smallest identifer among the colliding particles
can be used for that collision.  The third
approach can be used to sample exactly one random number per stream per
collision.  All approaches seem appropriate for reversal, depending on
the modeler's specific needs regarding computational cost and the
stream period.

Another context in which reversibility considerations of random number
streams plays an important role is in generating random samples
of ${\bm \Psi}$.  The \textbf{${\bm G}$-to-${\bm \Psi}$} function
for sampling ${\bm \Psi}$
involves sampling the circumference of an ellipse (or, in general,
sampling the points on the surface of higher dimensional ellipsoids).
While rejection-based sampling procedures \cite{Press2007}
are available for such problems, they cannot be used in reversible execution.
This is because the number of (uniformly distributed) random
numbers used by such rejection-based procedures
varies with each sampled point, which makes it
impossible to unambiguously reverse the random number stream without
keeping track of how many random numbers were generated for each sample.
In fact, rejection-based sampling can only be employed for certain
special classes of probability distributions, whose parametric input does
not vary across samples \cite{Perumalla2009}, whereas,
in sampling ${\bm \Psi}$, the parametric input varies with each collision.

\section{Implementation Results}\label{sec:results}

In order to test the performance of the
algorithms, we implemented the algorithms in software using the C++ programming
language, and executed simulation experiments.  The experiments are intended to
test (1) the ability to restore the initial state after a sequence of many
forward collisions followed by their reversals, and (2) the quality of phase
space coverage discerned from the uniformity of generated velocity samples.
For random number generation, we used a reversible version of a high quality
linear congruential generator \cite{LEcuyer1997} with a period of $2^{121}$.  A
single generator stream is used for all the particles.  In each configuration,
the experiments verify successful reversal across several thousands of
collisions.

\subsection{Collision Sequence Reversal}

\begin{algorithm}[h]
    \caption{\textit{Reversal Illustration for 2-Particle Collisions in 2 Dimensions}}
    \label{algo:impltest-2part-2dim}
    \begin{algorithmic}[1]
        \STATE $(a,b,c,d) \leftarrow (a_0,b_0,c_0,d_0)$
	    \COMMENT{initial velocities}
	\STATE $\alpha \leftarrow a+b$,
	       $\beta \leftarrow c+d$,
	       $\delta \leftarrow a^2+b^2+c^2+d^2$
	    \COMMENT{momenta and energy}
        \STATE $S \leftarrow $ random number seed
        \STATE $r_{21x} \leftarrow 1$,
               $r_{21y} \leftarrow 1$
	    \COMMENT{normalized collision geometry}

        \FOR[\textbf{$\---$ Forward Execution $\---$}]{$i=1$ to $N_c$}
	\STATE $\phi_1^{'} \leftarrow {\bm{V\text{-to-}\Phi}}(a,b,c,d)$ of
	        \refsec{2spheres-2dim}
	\STATE $G \leftarrow {\bm {RNG}}(S)$
	    \COMMENT{Generate next random number in $[0,1)$}
	\STATE $\phi_1 \leftarrow$ Apply \refalgo{2part2dimforward}
               on $\phi_1^{'}$ using $G$, $r_{21x}$, and $r_{21y}$
        \STATE $(a,b,c,d) \leftarrow {\bm{\Phi\text{-to-}V}}(\alpha,\beta,\delta,\phi_1)$ of \refsec{2spheres-2dim}
	\STATE $a\leftarrow-a$, $b\leftarrow-b$, $c\leftarrow-c$, $d\leftarrow-d$
	    \COMMENT{Reverse velocities to create next collision}
        \ENDFOR

        \FOR[\textbf{$\---$ Reverse Execution $\---$}]{$i=N_c$ to $1$}
	\STATE $a\leftarrow-a$, $b\leftarrow-b$, $c\leftarrow-c$, $d\leftarrow-d$
	\STATE $\phi_1 \leftarrow {\bm{V\text{-to-}\Phi}}(a,b,c,d)$ of
	        \refsec{2spheres-2dim}
	\STATE $G \leftarrow {\bm {RNG^{-1}}}(S)$
	    \COMMENT{Recover previous random number}
	\STATE $\phi_1^{'} \leftarrow$ Apply \refalgo{2part2dimreverse}
               on $\phi_1$ using $G$, $r_{21x}$, and $r_{21y}$
        \STATE $(a,b,c,d) \leftarrow {\bm{\Phi\text{-to-}V}}(\alpha,\beta,\delta,\phi_1^{'})$ of \refsec{2spheres-2dim}
        \ENDFOR

	\IF[\textbf{$\---$ Verification $\---$}]{$a=a_0$ \AND $b=b_0$ \AND $c=c_0$ \AND $d=d_0$}
	    \PRINT 'Passed'
	\ENDIF
    \end{algorithmic}
\end{algorithm}

\begin{algorithm}[h]
    \caption{\textit{Reversal Illustration for 3-Particle Collisions in 1 Dimension}}
    \label{algo:impltest-3part-1dim}
    \begin{algorithmic}[1]
        \STATE $(a,b,c) \leftarrow (a_0,b_0,c_0)$
	    \COMMENT{initial velocities}
	\STATE $\alpha \leftarrow a+b+c$,
	       $\delta \leftarrow a^2+b^2+c^2$
	    \COMMENT{momenta and energy}
        \STATE $S \leftarrow $ random number seed
        \FOR[\textbf{$\---$ Forward Execution $\---$}]{$i=1$ to $N_c$}
	\STATE $\phi_1^{'} \leftarrow {\bm{V\text{-to-}\Phi}}(a,b,c)$ of \refsec{3spheres-1dim}
	\STATE $G \leftarrow {\bm {RNG}}(S)$
	    \COMMENT{Generate next random number in $[0,1)$}
	\STATE $\phi_1 \leftarrow$ Apply \refalgo{3part1dimforward}
               on $\phi_1^{'}$ using $G$
        \STATE $(a,b,c) \leftarrow {\bm{\Phi\text{-to-}V}}(\alpha,\delta,\phi_1)$ of \refsec{3spheres-1dim}
	\STATE $a\leftarrow-a$, $b\leftarrow-b$, $c\leftarrow-c$
	    \COMMENT{Reverse velocities to create next collision}
        \ENDFOR

        \FOR[\textbf{$\---$ Reverse Execution $\---$}]{$i=N_c$ to $1$}
	\STATE $a\leftarrow-a$, $b\leftarrow-b$, $c\leftarrow-c$
	\STATE $\phi_1 \leftarrow {\bm{V\text{-to-}\Phi}}(a,b,c)$ of \refsec{3spheres-1dim}
	\STATE $G \leftarrow {\bm {RNG^{-1}}}(S)$
	    \COMMENT{Recover previous random number}
	\STATE $\phi_1^{'} \leftarrow$ Apply \refalgo{3part1dimreverse}
               on $\phi_1$ using $G$
        \STATE $(a,b,c) \leftarrow {\bm{\Phi\text{-to-}V}}(\alpha,\delta,\phi_1^{'})$ of \refsec{3spheres-1dim}
        \ENDFOR

	\IF[\textbf{$\---$ Verification $\---$}]{$a=a_0$ \AND $b=b_0$ \AND $c=c_0$}
	    \PRINT 'Passed'
	\ENDIF
    \end{algorithmic}
\end{algorithm}

\refalgo{impltest-2part-2dim} shows the pseudocode of the
experiment program for 2-particle collisions in 2 dimensions.
\refalgo{impltest-3part-1dim} shows the pseudocode of the experiment program
for 3-particle collisions in 1 dimension.  In both, the state of
the particles is initialized to any desired intial configuration, followed
by a sequence of $N_c$ applications of the forward collision operator.
${\bm {RNG}}(S)$ represents the generation of the next sample in $[0,1)$
from the random number stream (which also updates $S$ as side-effect), while
${\bm {RNG^{-1}}}(S)$ represents the reversal of the most previous
invocation to ${\bm {RNG}}$ and the recovery of the most recently
generated value from ${\bm {RNG}}$.
After each forward collision, post-collision velocities are multiplied
by $-1$ to give the new pre-collision velocities for the next collision.
Since the post-collision velocities are guaranteed to be divergent, their
opposites are guaranteed to give converging velocities that will result
in the next collision.  After all $N_c$ forward collisions, the
entire sequence is reversed by executing the reverse collision operator $N_c$
times.  Clearly, the reversal is successful if the final velocities after all
$N_c$ reversals recovers the initial velocities.  This condition is verified at
the end and the corresponding status message is printed.

All executions terminated successfully with a ``passed''
status, verifying the restoration of initial state after reversal of
$N_c$ collisions.  The unbiased and correct phase space coverage is tested
with $N_c$ up to $10^6$ by plotting the
post-collision angles against random angle offsets and pre-collision
velocities.

In the case of 3-particle collisions in 1 dimension, numerical integration to
compute the segment length of an ellipse was performed using the Simpson's
rule.  Also, care was needed to account for numerical precision issues when
using the numerically computed cosine function, which sometimes produces
numerical noise for angles within very small neighborhoods of multiples of
$\tfrac{\pi}{6}$ and $\tfrac{\pi}{2}$.  A tolerance of $\pm 10^{-8}$ around
zero was employed when determining whether any given cosine value can be
considered zero or positive.

\begin{figure}[htb]
  \centering
  \begin{tabular}{c}
  \input{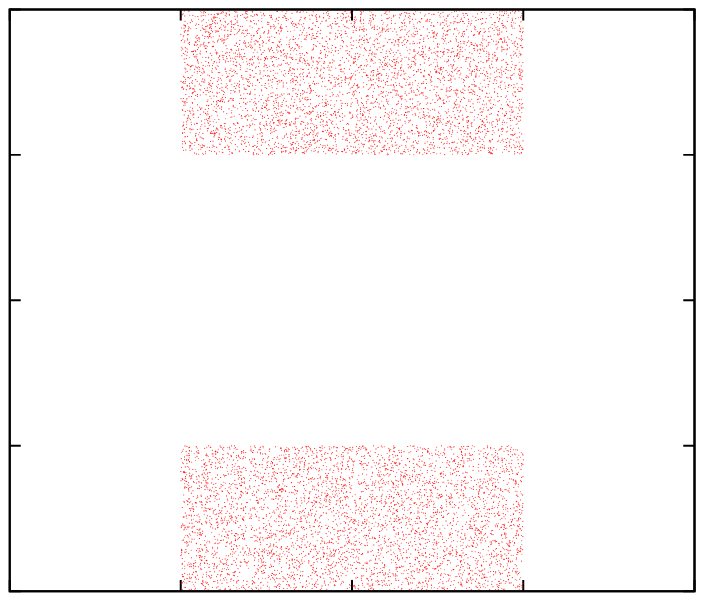}
      \\
  \input{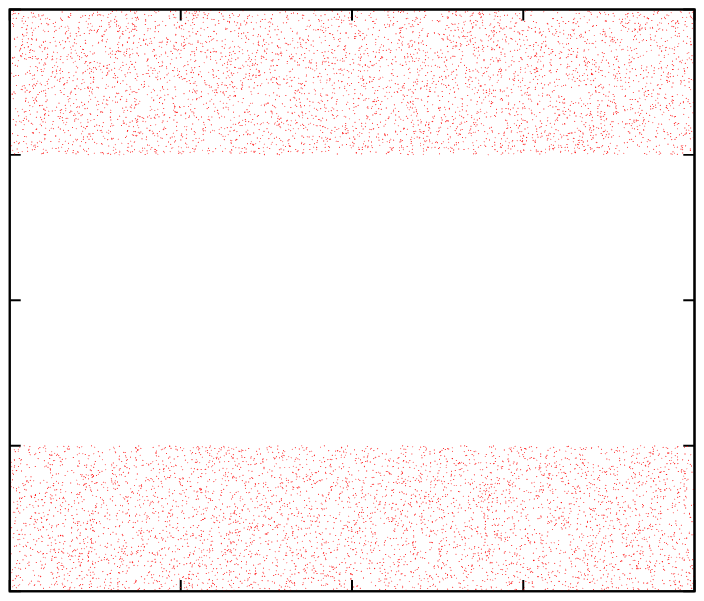}
  \end{tabular}
  \caption{Sampling for $N_c=10^4$ collisions in 2-particle collisions in 2 dimensions}
  \label{fig:2spheres-2dim-data-phi0}
\end{figure}

\begin{figure}[htb]
  \centering
  \begin{tabular}{c}
  \input{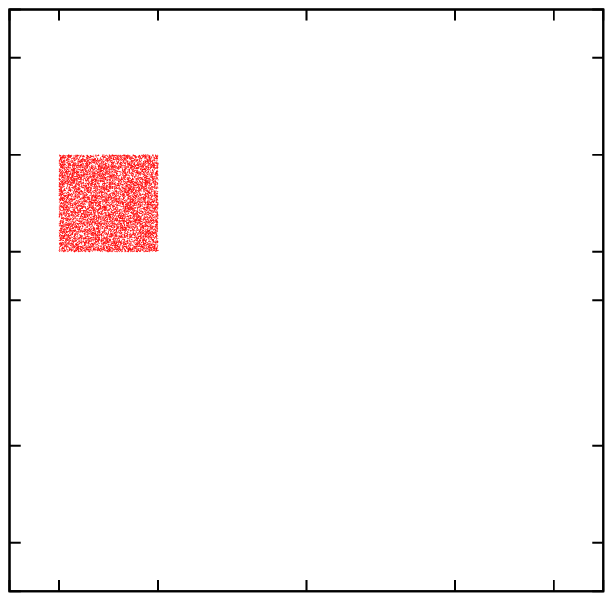}
      \\
  \input{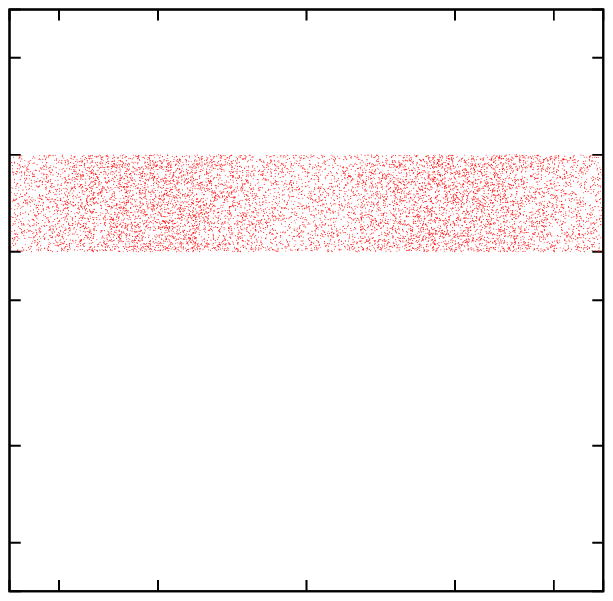}
  \end{tabular}
  \caption{Sampling for $N_c=10^4$ collisions in 3-particle collisions in 1 dimension}
  \label{fig:3spheres-1dim-data-phi0}
\end{figure}

\reffig{2spheres-2dim-data-phi0} plots%
\remove{ 
\footnote{The points in the plot may show banded patterns in print.  However,
such banding or other artifacts are due to printer characteristics
and are absent in the electronic form of the image.
}
} 
the post-collision angles against their
corresponding pre-collision angles generated as part of an execution of
2-particle collisions in 2 dimensions, and plots
the post-collision angles against their corresponding random angles in the same
execution.  Both plots show excellent uniformity and correctness of the
sampled phase space.  Similarly, \reffig{3spheres-1dim-data-phi0}
shows the corresponding data for 3-particle
collisions in 1 dimension; again, they display excellent coverage.  Since
the covered areas in the plots become too dense for visual clarity
when $N_c$ becomes large, only data for $N_c=10^4$ are shown in the figures
\footnote{Figures for much larger number of collisions ($N_c=10^5$) were also
generated confirming similar uniformity of coverage and correctness of sampled
phase space, but they result in much larger file sizes and hence not included here.}.

\subsection{Randomness Tests}
The correctness of phase space coverage is experimentally verified by testing
the randomness of the generated angles using statistical tests.
The Diehard Battery of Tests \cite{Diehard1995} for statistical verification of
randomness is used for this purpose.

The post-collision angles are mapped to double-precision numbers, each in
$[0,1)$, and the uniformity of their distribution is verified.  This method is
applied to 2-particle collisions in 2 dimensions, and to 3-particle collisions
in 1 dimension.

For 2-particle collisions in 2 dimensions, each post-collision angle
$\phi_1$ is mapped to a double-precision number $\eta\in[0,1)$ as
$\eta={(\phi_1-\kappa)}/{\pi}$, where $\kappa=0$ if $0\leq\phi_1\leq \tfrac{\pi}{2}$,
and $\kappa=\pi$ otherwise ($\tfrac{3\pi}{2}\leq\phi_1\leq2\pi$).
Each $\eta$ is converted into an integer equal to $\lfloor{\eta\times
4,294,967,296}\rfloor$, and the resulting series of numbers (in hexadecimal
format) is converted via the \texttt{asc2bin} program of Diehard to a
binary-formatted file given as input to the \texttt{diehard} program.

For 3-particle collisions in 1 dimension, each post-collision angle
$\phi_1$ is mapped to a double-precision number $\eta\in[0,1)$ as
$\eta={(\phi_1-\tfrac{7\pi}{6})}/{(\tfrac{3\pi}{2}-\tfrac{7\pi}{6})}$.  Each
$\eta$ is converted into an integer equal to $\lfloor{\eta\times
4,294,967,296}\rfloor$, and the resulting series of numbers (in hexadecimal
format) is converted via the \texttt{asc2bin} program of Diehard to a
binary-formatted file given as input to the \texttt{diehard} program.

Both \refalgo{impltest-2part-2dim} and \refalgo{impltest-3part-1dim} were
successfully executed with
$N_c=3,000,000$ collisions, such that they terminate with a
``Passed'' result.  The angles are logged to a file during their execution,
and then used as input to the randomness tests.  According to the Diehard
tests, if the ``$p$-values'' computed and printed by \texttt{diehard}
are observed to be strictly greater than 0 and less than unity,
randomness is understood to be satisfied \cite{Diehard1995}.
The $p$-values observed from the randomness tests on the
generated angles are given in
\reftab{diehard-pvalues}.  Good phase space coverage via randomization
is indicated by the fact that the $p$-values from several tests are
significantly away from zero and unity.

For 2-particle collisions in 2 dimensions, all tests in the Diehard repository
were used and verified to generate very good randomness (positive $p$-values
less than unity) without exception.  This is due to the fact that no numerical
precision effects are present in the collision algorithm for 2-particle
collisions in 2 dimensions.  For 3-particle collisions in 1 dimension, all
tests were used except those that operate on bit-level representations (such as
the Birthday, Bitstream and Count-the-1s tests).  This is because round off and
trunction effects in numerical integration, even at a relatively high precision
of $10^{-9}$ in the computation of angles, introduces non-random patterns in
the last few bits of the mantissa, appearing as non-randomness when
selectively viewed in isolation or across multiple floating point
numbers.  However, when the angles are viewed as numbers themselves, uniform
randomness is indeed observed, as expected.

\afterpage{\clearpage}

\begin{table}[tbph]
  \caption{Randomness indicator $p$-values from Diehard battery of tests}
  \label{tab:diehard-pvalues}
  \centering
  \small
  \begin{tabular}{|l|l|r|r|} \hline
    \textbf{Test} & \textbf{Category} & \multicolumn{2}{|c|}{\textbf{$p$-value}}\\\hline
     &  & 2-Dimension & 1-Dimension \\
     &  & 2-Particle & 3-Particle \\\hline\hline
      CRAPS      & Overall                     & 0.979800 & 0.570270 \\
                 & Wins                        & 0.790715 & 0.862293 \\
                 & Throws/game                 & 0.979797 & 0.570273 \\\hline
      RUNS       & Set 1 - Up                  & 0.847070 & 0.298998 \\
                 & Set 1 - Down                & 0.128221 & 0.632054 \\
                 & Set 2 - Up                  & 0.183069 & 0.871807 \\
                 & Set 2 - Down                & 0.244909 & 0.111657 \\\hline
      SUMS       & 10 $\chi^2$-tests on 100 $\chi^2$-tests & 0.370145 & 0.257959 \\\hline
      SQUEEZE    & 42 degrees of freedom       & 0.890107 & 0.435410 \\\hline
      3DSPHERES  & $\chi^2$-test on 20 $p$-values & 0.813199 & 0.897844 \\\hline
      MINDIST    & $\chi^2$-test on 100 min-distances & 0.348209 & 0.751603\\\hline
      PARKLOT    & $\chi^2$-test on 10 $p$-values & 0.947200 & 0.720233 \\\hline
  \end{tabular}
\end{table}

\section{Performance Estimation}\label{sec:performance}

To estimate the performance gain that can be expected by resorting to reversible
collisions instead of state saving, we implemented a simulation of a sequence
of 2-particle collisions in a closed system of $N$ particles in $d=2$ dimensions.
Experiments were run with $N=1,000$, $N=10,000$, and $N=100,000$ particles.
Starting with random initial configuration of positions and velocities, and randomly
selected inter-collision times, particle motion is simulated between collisions,
and, at every collision point, the collision operator
is applied on a random pair of particles.  With state saving, the
system state is saved to memory before every collision, to be able to
roll back to that state.  With reverse computation, no state is saved,
as the system can be rolled back perfectly to any point in the past
by reverse computation alone, without reliance on memory.

Two platforms with different computational and memory
characteristics are tested: one with traditional central
processing unit (CPU), and the other with newer graphical processing unit (GPU).
Modern CPUs now have much higher computational speeds than memory speeds;
the differential between computational and memory speeds is even more pronounced
in modern GPU platforms \cite{Pharr2005}.
Implementation on the CPU is realized in the C++ programming language, and that
on the GPU is in the CUDA programming language \cite{Sanders2010}.
The CPU is an AMD Opteron 6174 processor with 64 GB of memory.  The GPU
is a high-end nVidia Geforce GTX 580 (Fermi) accelerator
with 512 CUDA cores and 3 GB device memory.  Compilation systems used were
gcc 4.4.5 and CUDA 4.1.

In each simulation run, $N_c=1000$ collisions were simulated, and, after
$N_c$ collisions, the system was rolled back to the beginning.
With reverse computation, the positions and velocities are verified
to match the initial conditions exactly (to within at least $\epsilon=\pm 10^{-9}$),
while, with state saving, the results are trivially exactly matched.

\begin{figure}[htbp]
  \centering
  \begin{tabular}{|c||c|}\hline
      CPU-based implementation & GPU-based implementation
          \\
          \hline
          \hline
      \multicolumn{2}{|c|}{(a)~~~~~~~~~~~~~~Number of particles = 1,000~~~~~~~~~~~~~~(b)}
          \\
      \includegraphics[width=0.45\textwidth,keepaspectratio]{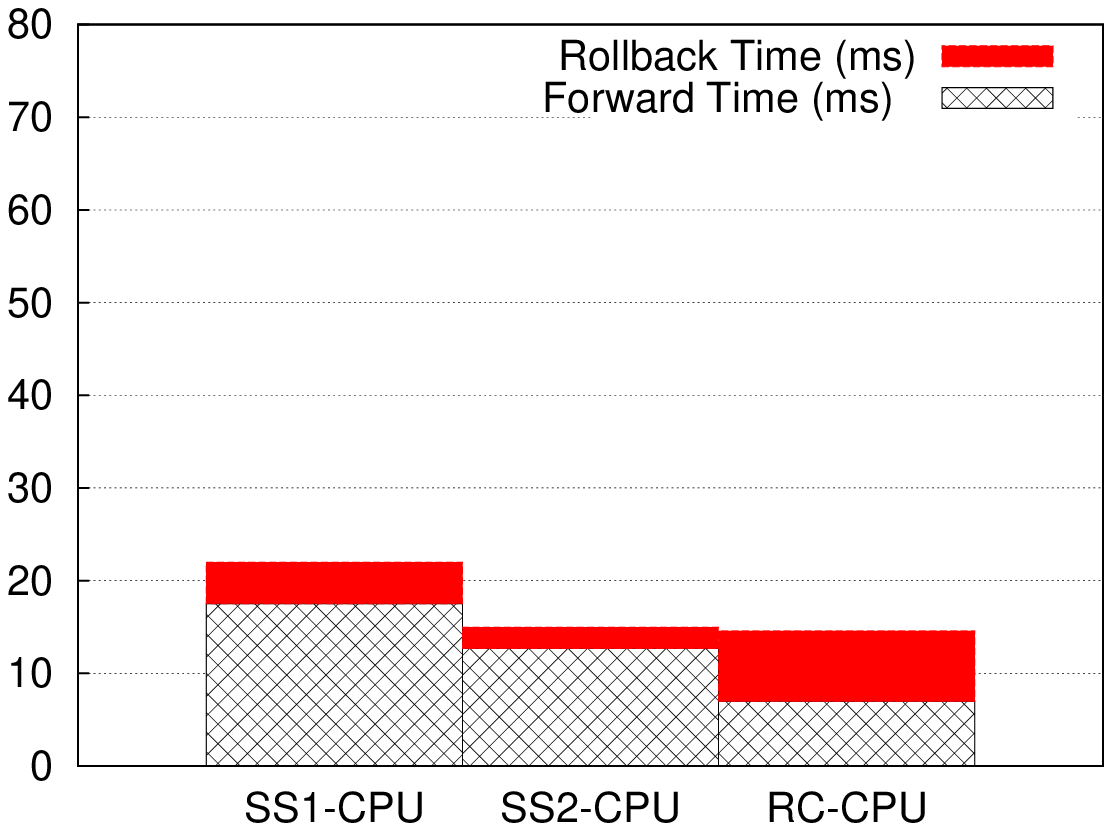}
          &
      \includegraphics[width=0.45\textwidth,keepaspectratio]{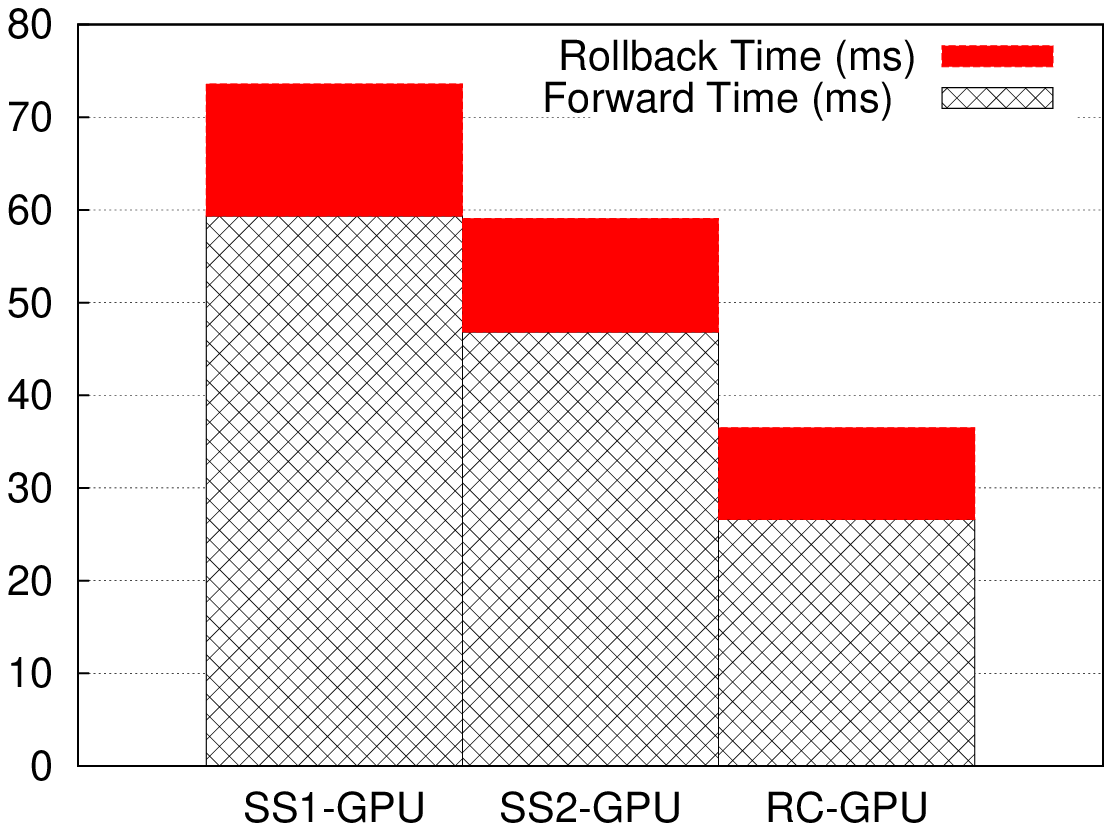}
          \\
          \hline
          \hline
      \multicolumn{2}{|c|}{(c)~~~~~~~~~~~~~~Number of particles = 10,000~~~~~~~~~~~~~~(d)}
          \\
      \includegraphics[width=0.45\textwidth,keepaspectratio]{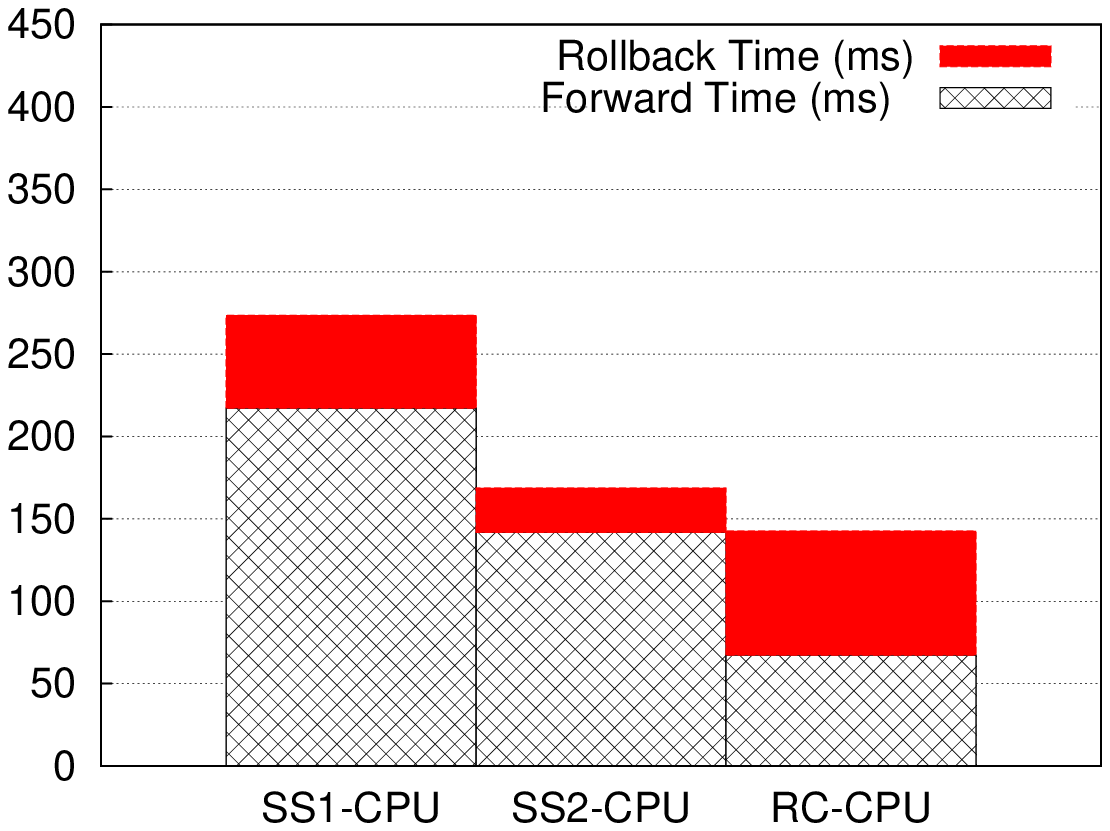}
          &
      \includegraphics[width=0.45\textwidth,keepaspectratio]{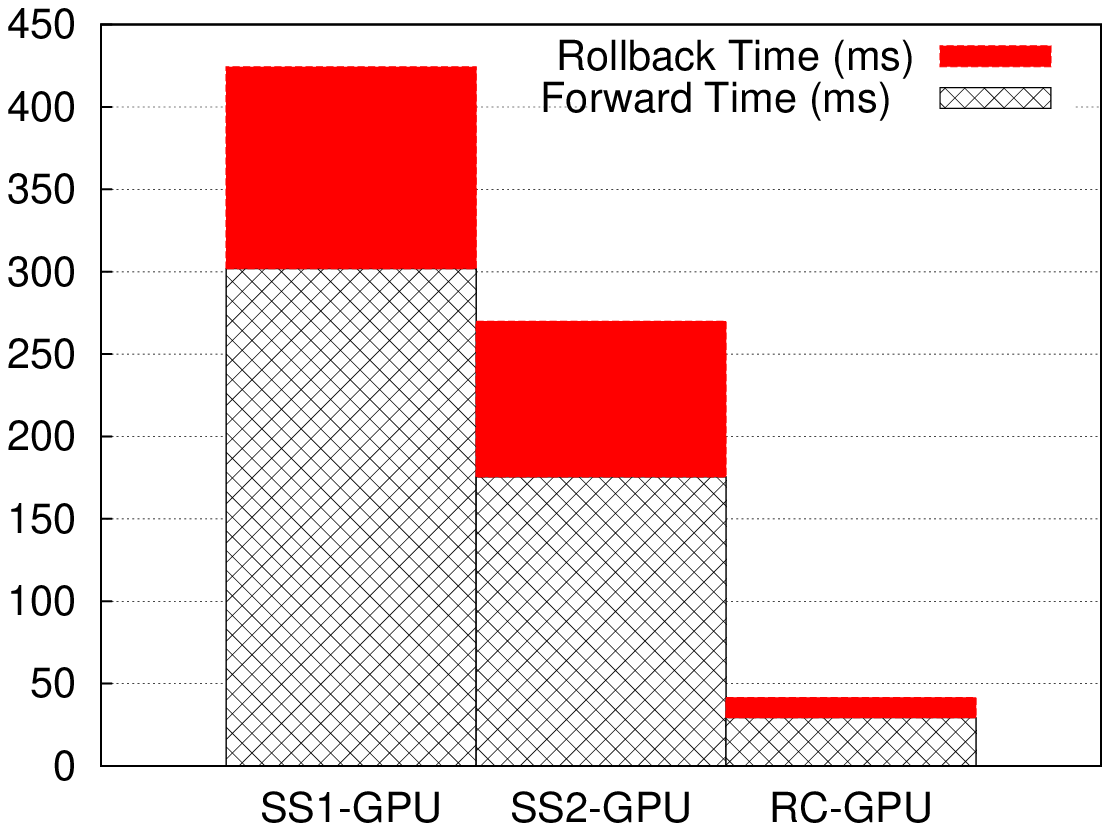}
          \\
          \hline
          \hline
      \multicolumn{2}{|c|}{(e)~~~~~~~~~~~~~~Number of particles = 100,000~~~~~~~~~~~~~~(f)}
          \\
      \includegraphics[width=0.45\textwidth,keepaspectratio]{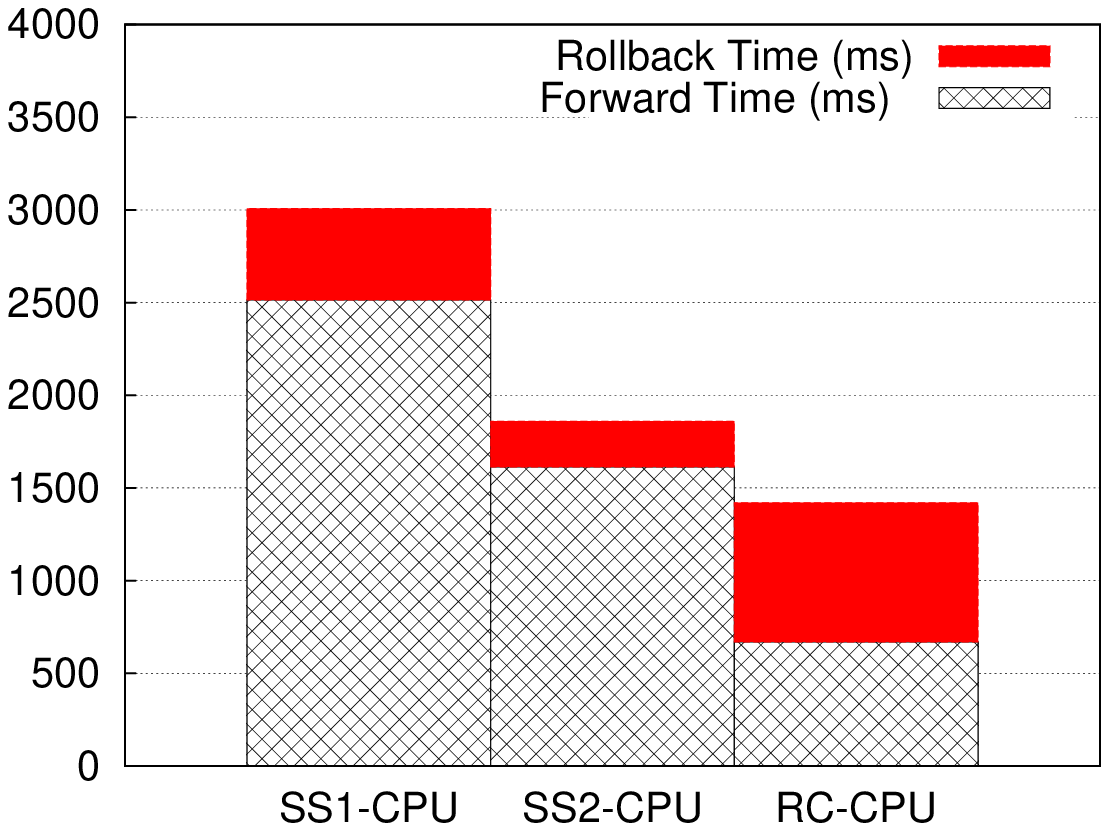}
          &
      \includegraphics[width=0.45\textwidth,keepaspectratio]{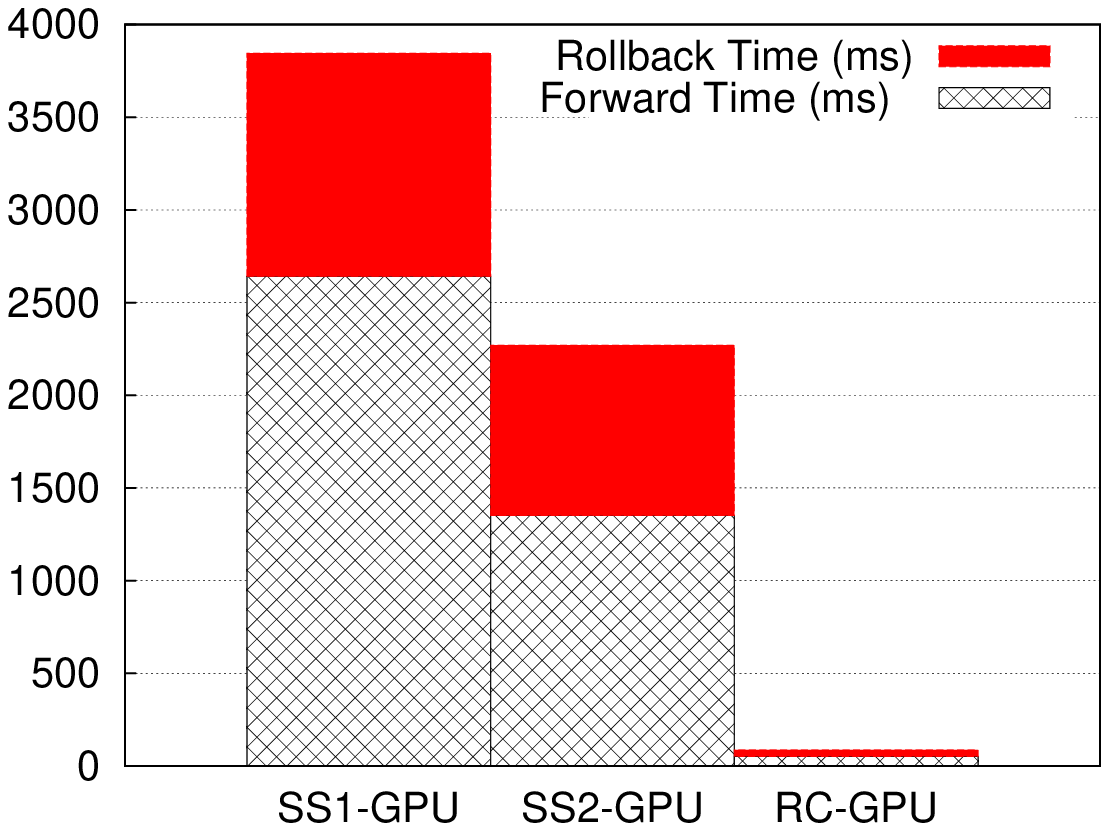}
          \\
          \hline
  \end{tabular}
  \caption{Performance of state saving and reverse computation with $N_c=1000$ collisions}
  \label{fig:performance}
\end{figure}

The results are drawn as stacked histograms in \reffig{performance}, with
the total height of each bar representing the total time for forward execution
of $N_c$ collisions and their reversal, which is split into \textit{Forward
Time} and \textit{Rollback Time} in milliseconds.  Three variants are
benchmarked: \textbf{SS1} represents the state saving mechanism in which
all state is saved (positions and velocities); \textbf{SS2} represents an
optimized state saving mechanism in which the positions are saved and
only the four components of the pre-collision velocities of the colliding pair
are saved; \textbf{RC} represents the reverse computation with no memory.
For each variant, the suffix \textbf{-CPU} represents execution on the CPU,
and \textbf{-GPU} represents execution on the GPU.

Total run time with reverse computation is lower across the board.
As expected, the greatest differential is observed in
\reffig{performance}(f) for the GPU runs with the
largest number of particles ($N=100,000$).  Since the ratio of memory transfer
cost to computational cost is much higher with the GPU, reverse computation
runs much faster, while state saving incurs the high memory transfer cost for
every collision. Also, the GPU platform is known to be extremely efficient with
large vectorized codes such as this simulation (i.e., more particles can be 
simulated with little increase in total time, if memory bottle neck is relieved).
Hence, the reverse computation runs are extremely fast on the GPU,
compared to all CPU runs and
also compared to state saving with GPU.  On smaller number of particles ($N=1000$
and $N=10,000$), CPU runs in (a) and (c) are faster than GPU runs in (b) and (d)
because of larger CPU caches, yet, even in this case
of relatively lower memory cost, reverse computation is observed to run faster.
Even more importantly, when the number of particles is further increased (e.g.,
$N\geq 1\text{ million}$), state saving becomes infeasible due to memory
limitations, but reverse computation runs well even at such large scale.
Similarly, the benefit of using reversible simulation only increases with
increase in the number of collisions, due to corresponding increase
in the memory needs of state saving.
A more detailed analysis of the memory subsystem behavior (e.g., data cache
misses at levels 1 and 2, and translation lookaside buffer metrics) for each
of these runs is part of our planned future work.

\section{Summary}\label{sec:summary}

The classical problem of simulating elastic collisions of
hard spheres has been revisited, with the important additional requirement of
reversibility.  Although classical simulation of elastic collisions
has been well studied in the literature, little has been known on how to
simulate them reversibly with minimal memory overhead.
Here, we formalized the problem in terms of accurate phase space coverage
specification and geometrical constraints.  We solved the problem by
developing a general framework that combines reversible pseudo random
number generation with new mapping functions, geometrical
constraints, and reversal semantics.
While previous log-based approaches require
memory proportional to the number of collisions, our algorithms incur
essentially zero memory overheads and also ensure correct phase space coverage.
\remove{%
The key insight in developing our approach is that
we define the post-collision angles in terms of reversible random offsets
(sampled from all permissible phase space) from pre-collision angles.
} 
We developed the detailed steps for 2-particle collisions (up to 3 dimensions)
and 3-particle collisions (up to 2 dimensions).
In these configurations, memory overhead is exactly zero for collisions in
which $d_n=1$, and essentially zero for collisions with $d_n>1$.  In the latter
configurations, $\{\phi_{i+1}^{'},\ldots,\phi_{d_n}^{'}\}$ are logged if and
only if $\phi_{i}=0$, for any $1\leq i < d_n$.
Generalizations to collisions among
larger number of particles and at higher dimensions are tedious, but can be
carried out if needed.  At higher dimensions and with larger number of
particles, computationally expensive numerical integration becomes necessary in
both classical (forward-only) approaches as well as in the \textit{forward}
procedures of our reversible method.  To meet the goal of minimal
memory overhead, our \textit{reverse} procedures rely on numerical integration
as well, whereas log-based reversal approaches would use memory to save
information and avoid recomputation in the reverse path.  In a normal,
well-balanced parallel execution, reversals of collisions are far fewer than
forward collisions (i.e., dependency violations are incurred infrequently).  In
such cases, our reversible models are more efficient, since forward cost for
reversibility is eliminated, whereas log-based approaches would incur
log-generation costs in forward execution.
\remove{%
However, if the number of reversals
is large (e.g., processors are ill-balanced and incur many ordering
violations), the computation cost for our reverse procedures is larger than the
memory restoration cost of log-based procedures.  Thus, there is a trade-off
between memory and computation for reversibility in certain configurations,
which remains to be explored.
} 

Finally, although reversibility of elastic collisions is a seemingly simple
problem to formulate, it includes sufficient complexity to make it
challenging.  The new approach and results presented here
offer insights in revisiting additional classical physical system models with
the added requirement of reversibility and exploring their fundamental memory
characteristics and limits.

\bibliographystyle{alpha}
\bibliography{rmd}

\remove{
\section{Discussion}\label{sec:discussion}
\subsection{Significance of Geometrical Conditions in Reversibility}

Illustrate why disregarding the geometrical constraints introduces
ambiguities on backward/reversal path, defeating perfect reversibility.

\subsection{Potential Bias in Phase Space Sampling in Reversibility}

Illustrate the issue of bias for incorrect solutions that can be reversed
without necessarily sampling the full phase space without bias.
}


\section*{Acknowledgements}

This paper has been authored by UT-Battelle, LLC, under contract
DE-AC05-00OR22725 with the U.S. Department of Energy. Accordingly, the
United States Government retains and the publisher, by accepting the
article for publication, acknowledges that the United States Government
retains a non-exclusive, paid-up, irrevocable, world-wide license to
publish or reproduce the published form of this manuscript, or allow others
to do so, for United States Government purposes. Research was partly supported
by the DOE Early Career Award in Advanced Scientific Computing Research
under grant number 3ERKJR12.
The authors thank Alfred J. Park, Sudip K. Seal, and James J. Nutaro for
constructive comments on early versions of the manuscript, and Alfred J. Park
for helping with the implementation for performance estimation.


\end{document}